\documentclass[aps,prb,11pt]{revtex4-1}
\usepackage{bm}
\usepackage[dvips]{graphicx}
\usepackage{bm}
\begin{document}
\title{Towards a Microscopic Theory of the Knight Shift in an Anisotropic, Multiband Type-II Superconductor}


\author{Richard A. Klemm}


\affiliation{Department of Physics, 4111 Libra Drive, University of Central Florida, Orlando, Florida 32816-2385 USA}




\date{\today}

\begin{abstract}
A method is proposed to extend the zero-temperature Hall-Klemm microscopic theory of the Knight shift $K$ in an anisotropic and correlated, multi-band metal to calculate $K(T)$ at finite temperatures $T$ both above and into its superconducting state. The transverse part of the magnetic induction ${\bf B}(t)={\bf B}_0+{\bf B}_1(t)$  causes adiabatic changes suitable for treatment with the Keldysh contour formalism and analytic continuation onto the real axis.  We propose that the Keldysh-modified version of the Gor'kov method can be used to evaluate $K(T)$ at high ${\bf B}_0$ both in the normal state, and by quantizing the conduction electrons or holes with Landau orbits arising from ${\bf B}_0$, also in the entire superconducting regime for an anisotropic, multiband Type-II BCS superconductor. Although the details have not yet been calculated in detail, it appears that this approach could lead to the simple result $K_S(T)\approx a({\bf B}_0)-b({\bf B}_0)|\Delta({\bf B}_0,T)|^2$, where  $2|\Delta({\bf B}_0,T)|$ is the  effective superconducting gap.  More generally, this approach can lead to analytic expressions for $K_S(T)$ for anisotropic, multiband Type-II superconductors of various orbital symmetries that could aid in the interpretation of experimental data on  unconventional superconductors.
\end{abstract}

\maketitle

\section{Introduction}
In nuclear magnetic resonance (NMR) measurements of a nucleus, there is a difference between the resonance frequency of the nucleus when it is in a metal from when it is vacuum or in an insulator.  This is known as the Knight shift \cite{Knight1949}.
Although the temperature $T$ dependence of the Knight shift in a superconductor has long been considered to be a probe of the spin state of the paired electrons \cite{Knight1956,Reif1956,Reif1957}, the only theoretical basis for the experiments was the 1958 assumption of Yosida that the probed nuclear spins could be entirely neglected \cite{Yosida1958}, and the only quantity of interest was the temperature $T$ dependence of the zero-field limit of the electron spin susceptibility of an isotropic and uncorrelated Type-I superconductor \cite{Yosida1958}. This led for a BCS singlet-pair-spin superconductor to a $T$ dependence of the Knight shift $K_S(T)$ proportional to $x/(1+x)$, where $x=(\beta/\Delta)\frac{d\Delta}{d\beta}$,  $\beta=1/T$, where we set $k_B=1$, and $\Delta$ is one-half the BCS energy gap, for which $K_S(T)\rightarrow0$ as $T\rightarrow0$, unlike most experimental results \cite{Yosida1958}.

 For isotropic Type-I superconductors in the Meissner state, crushing the sample to a powder of crystallites the cross-sections of which were less than the magnetic penetration depth was usually found to provide a reasonable method for that conventional theory to be applicable \cite{Knight1956,Reif1956,Reif1957}.  In the first years following the BCS theory, the transition metal superconductors were found to behave somewhat differently, as the Knight shift did not vanish as $T\rightarrow0$ \cite{Gladstone1969}, and it was thought that surface spin-orbit scattering could explain the near-cancelation of the Knight shift in transition metals \cite{Gladstone1969,Abrikosov1962}.  But surface impurity spin-orbit scattering could not explain the observed non-vanishing $K(0)$ results observed  in clean materials. It is now understood that there is also a component to the Knight shift due to the orbital motion of the electrons in a superconductor, and for an anisotropic superconductor, this orbital contribution to the Knight shift depends upon the  magnetic induction ${\bm B}$ direction.

 There have since been many examples of unexplained behaviors of the Knight shift in exotic superconductors.  Since one possibility of a $T$-independent Knight shift result would be a parallel-spin, triplet pair-spin superconducting state, the use of the Knight shift has been considered to be a principle tool for the identification of a triplet pair-spin state.  Some examples of triplet-pair-spin or some other types of exotic behavior claimed to exist in unusual materials based upon the unconventional Knight shift $T$-dependence are listed in the bibliography \cite{Baek2013,Kohori2001,Lee2002,Lee2003,Michioka2006,Kato2006,Sakurai2015}.  But one of those materials was a quasi-one-dimensional organic superconductor \cite{Lee2002,Lee2003}, some examples of which often exhibit spin-density waves \cite{Klemm2000}, and another was the very dirty sodium cobaltate hydrate material \cite{Michioka2006,Kato2006,Sakurai2015}.  In the latter example, the upper critical field parallel to that layered compound is Pauli-limited, which normally only occurs when the magnetic field breaks the oppositely-oriented pair spins \cite{Chou2004,Klemm2012}. Since dirt drastically suppresses $p$-wave superconductivity \cite{Scharnberg1980}, the sodium cobaltate hydrate Knight shift results, if correct, are likely to arise from some other mechanism.

 Moreover, in highly anisotropic Type-II superconductors, such as the cuprates and heavy fermion materials, other significant breakdowns in the Yosida theory have been found to exist. In the first Knight shift measurements on
 the cuprate YBa$_2$Cu$_3$O$_{7-\delta}$, Bennett {\it et al.} found that although the Yosida theory appeared to work for the $^{63}$Cu spins in the CuO chains for all field directions, although the orbital contributions are different for each of the three orthogonal applied field directions, and it also appeared to work well for the $^{63}$Cu spins in the CuO$_2$ planes when the strong constant magnetic field ${\bm H}$ was applied parallel to the CuO$_2$ layers.  But, when ${\bm H}$ was applied normal to the CuO$_2$ layers, no $T$ dependence to the Knight shift was observed in that cuprate\cite{Barrett1990}.  This result was later described by Slichter as possibly being due to a ``fortuitous'' cancelation of the effect from an isolated planar $^{63}$Cu spin  by its interaction with its near-neighbor planar $^{63}$Cu spins \cite{Slichter1999}.  Subsequently, in a number of layered correlated superconductors, the $T$ dependence of the Knight shift probes of the nuclear spins in the layers with the field applied normal to the layers has been observed to vary strongly with field strength, approaching a constant $K_S(T)$ in the large normal field strength limit, as first observed by Bennett {\it et al.} \cite{Barrett1990,Slichter1999,Fujiwara1991,Zheng1996,Zheng2003,Kotegawa2008}.

Especially in the case of Sr$_2$RuO$_4$, numerous Knight shift measurements of the $^{17}$O, $^{99}$Ru, $^{101}$Ru, and $^{87}$Sr have all led to temperature-independent Knight shift measurements \cite{Ishida1998,Ishida2001,Mackenzie2003}, as did polarized neutron scattering experiments \cite{Duffy2000}.  This experiments were all interpreted as evidence for a parallel-spin pair state in that material. However, several upper critical field measurements with the field parallel to the layers showed strong Pauli limiting effects \cite{Deguchi2002,Kittaka2009}, which is inconsistent with a parallel-spin pair state\cite{Machida2008,Zhang2014a}. In addition, the fact that $T$-independent Knight shift measurements were obtained for the field both parallel and perpendicular to the RuO$_2$ layers is incompatible with any of the crystal point-group-compatible $p$-wave states.  Thus, the only way for a $T$-independent Knight shift to legitimately arise from a parallel-pair-spin state in both field directions is for the ${\bm d}$-vector (the vector describing the components of the three triplet spin states) to rotate with the magnetic field \cite{Annett2007,Leggett1975}.  This argument was used to show that while the upper critical field of Sr$_2$RuO$_4$ is strongly Pauli limited for the field applied parallel to the layers, it could possibly be consistent with one or more $p$-wave helical states, provided that the ${\bm d}$-vector is allowed to rotate freely with the magnetic field direction \cite{Zhang2014a}.  This means that spin-orbit coupling with the lattice would have to be negligible.  However, there is strong evidence that spin-orbit coupling in Sr$_2$RuO$_4$ is very strong at some points on the Fermi surface, ruling out such ${\bm d}$-vector rotation possibilities\cite{Rozbicki2011}.  More worrisome for the Knight shift measurement results is the fact that carefully performed scanning tunneling measurements of the electronic density of states provided very strong evidence of a nodeless superconducting order parameter orbital symmetry in Sr$_2$RuO$_4$ \cite{Suderow2009,Klemm2012}, consistent with a nearly isotropic gap function that is essentially identical on all three of its Fermi surfaces. Since the theories behind the Pauli limiting effects and the BCS gap density of states are very well established, but the Knight shift measurement interpretations rely entirely on the complete neglect of the probed nuclear spins, the development of a microscopic theory of the $T$ dependence of the Knight shift in anisotropic and correlated Type-II superconductors is sorely needed.

We further note that the time dependence of a spin-1/2 particle in a classic magnetic resonance experiment is now a textbook example of an exactly soluble first quantization quantum mechanics problem giving rise to a Berry phase \cite{Berry1984,Griffiths2005}. In that case, the Berry, or geometric, phase is a combination of the resonance profile with the frequency of the oscillatory transverse applied magnetic field.  In higher spin $I$ systems, there are $2I$ combinations of those two quantities, giving rise to a multiplet of Berry phases, as discussed in the following.   Note that the probed nuclear spins of Sr$_2$RuO$_4$ are either 5/2 or 9/2.  Since nothing was known about the Berry phase in 1958, its possible implications for the  interpretation of Knight shift measurements have been generally and perhaps completely ignored in the literature.

In fairness to the pioneering work of Yosida \cite{Yosida1958}, there have been a few cases in which a complete lack of any $T$-dependence to the Knight shift has been confirmed by other experiments consistent with  a parallel-pair-spin superconducting state \cite{Hattori2013,Aoki2012,Gannon2012}.  These are for the uranium-based compounds UCoGe and UPt$_3$, for which the $T$-independent Knight shift in UCoGe is in agreement with the general assessment of the upper critical field and muon depolarization experiments \cite{Scharnberg1980,Aoki2012}.  In UPt$_3$, the seeming incompatibility of the Knight shift and the upper critical field appears to have been resolved by polarized neutron diffraction experiments\cite{Gannon2012}, favoring a parallel-spin pair state in all three superconducting phases.  In the ferromagnetic superconductors UGe$_2$, UCoGe, and URhGe, the weak Ising-like ferromagnetism appears to allow for a parallel-spin, $p$-wave superconducting order parameter in the plane perpendicular to the ferromagnetism, but the Knight shift measurements have not yet been made on URhGe and UGe$_2$, the latter of which is only superconducting under pressure. In these three ferromagnetic superconductors, there is at least a plausible mechanism for a parallel-spin pair superconducting state, and in URhGe the upper critical field fits the predictions for all three crystal axis directions of a parallel-spin $p$-wave polar state fixed to the crystal $a$-axis direction normal to the $c$-axis Ising ferromagnetic order \cite{Aoki2012,Scharnberg1985}, and there is a reentrant, high field phase that violates the Pauli limit by a factor of 20\cite{Aoki2012}. In order to obtain further evidence that the classic Yosida interpretation of a $T$-independent $K_S(T)$ can correctly imply a parallel-spin superconducting state, we urge that $^{73}$Ge, with a strong nuclear moment, (or possibly $^{103}$Rh, with a much weaker nuclear moment) $K_S(T)$ measurements on URhGe be carefully performed in the low-field superconducting phase.

\section{The Model}
The first microscopic model of  the Knight shift at $T=0$ in anisotropic and correlated metals was recently presented by Hall and Klemm \cite{Hall2016}.  This model assumed that the applied magnetic fields probe the nuclear spins, and the spins of the electrons orbiting the nucleus interact with the nucleus via the hyperfine interaction in the form of a diagonal ${\bm g}$ tensor with two distinct components $D_{x}=D_y\ne D_z$. The assumption $D_x=D_y$ was made to simplify the calculations, as discussed in more detail in the following.  After interacting with the nuclear spins, the orbital electrons can be excited into one of multiple bands, each of which was assumed to have an ellipsoidal Fermi surface of arbitrary anisotropy and shape.  The orbital motion of the electrons in each of these bands was constrained by the  strong, time-independent part ${\bm B}_0$ of the magnetic induction ${\bm B}(t)$ to be in Landau levels, and the electron spins also could interact weakly with ${\bm B}_0$.  It was found that the self-energy due to $D_z$ led to the Knight shift, and that due to $D_x=D_y$ led to the first formulas for the linewidth changes associated with the Knight shift at $T=0$. However, since those calculations were made at $T=0$, they could not be used to probe the superconducting state.  In the following, a method is proposed to do so.

Following Haug and Jauho \cite{Haug2008}, we write  the Hamiltonian as ${\cal H}={\cal H}_0 +{\cal H}_{\rm int}+{\cal H}'(t)$, where ${\cal H}_0+{\cal H}_{\rm int}$ is the time-independent part and ${\cal H}'(t)$ is the time-dependent part due to the oscillatory (or pulsed) magnetic field transverse to the constant applied magnetic field ${\bm H}_0$, and the time-independent part  consists of the simple (or exactly soluble) part ${\cal H}_0$ and the interaction part ${\cal H}_{\rm int}$ that involves the interactions between the particles that must be treated perturbatively.  In the case at hand, there are four types of particles:  (1) the nuclear spins probed in the NMR experiment, which are assumed to have the general spin $I\ne0$ with $(2I+1)$ substates denoted $m_I$, (2) the local orbital electrons surrounding each of the nuclei probed in the NMR experiment, (3) the conduction electrons or holes that propagate from the local nuclei throughout the metal/superconductor, and (4) the superconducting Cooper pairs of electrons or holes.  We note that complicated materials such as Sr$_2$RuO$_4$ contain multiple Fermi surfaces, which can be a mix of electron and hole Fermi surfaces.  In this model, we do not account for competing ferromagnetism or charge-density wave (CDW) or spin-density wave (SDW) formation, at least one of which is normally present in the transition metal dichalcogenides, the organic layered superconductors, the cuprates, the iron pnictides, and the ferromagnetic superconductors.  Such competing effects will be the subjects of future studies.
\subsection{The Simple Hamiltonian ${\cal H}_0$}
Since in an NMR experiment,  the applied magnetic field can be applied in any direction, we assume the resulting constant magnetic induction ${\bm B}_0=B_0(\sin\theta\cos\phi,\sin\theta\sin\phi,\cos\theta)=B_0\hat{\bm r}$ with respect to the crystalline Cartesian $x,y,z$ axes.  We then quantize the spins along ${\bm B}_0$.
We thus write
\begin{eqnarray}
{\cal H}_0&=&{\cal H}_{{\rm n},0}+{\cal H}_{{\rm e},0} +{\cal H}_{{\rm cond},0}\hskip10pt,\label{eqn1}
\end{eqnarray}
where
\begin{eqnarray}
{\cal H}_{{\rm n},0}&=&-\omega_{\rm n}\sum_{i,m_I}m_Ia^{\dag}_{i,m_I}a^{}_{i,m_I}\hskip10pt,\\
{\cal H}_{{\rm e},0}&=&\sum_{i,q,\sigma}[\epsilon_q-\sigma\omega_{\rm e}/2]b^{\dag}_{i,q,\sigma}b^{}_{i,q,\sigma}\hskip10pt,\\
{\cal H}_{{\rm cond},0}&=&\sum_{j,\sigma}\int d^3{\bm r}_j\psi^{\dag}_{j,\sigma}({\bm r}_j)\biggl(\sum_{\nu=1}^3\frac{1}{2m_{j,\nu}}[\nabla_{j,\nu}/{\rm i}-{\rm e}A_{j,\nu}({\bm r}_j)]^2-\sigma\omega'_{j,{\rm e}}/2\biggr)\psi^{}_{j,\sigma}({\bm r}_j)\hskip10pt,
\end{eqnarray}
where ${\rm e}$ is the electronic charge, $a^{\dag}_{i,m_I}$ creates a nucleus at the atomic position $i$ of spin $I$ in the subspin state $m_I=-I, -I+1,\ldots,I-1,I$, $b^{\dag}_{i,q,\sigma}$ creates an electron orbiting that nucleus at site $i$ with energy $\epsilon_q$ and spin-1/2 eigenstates indexed by $\sigma=\pm1$, where  $q\in(n,\ell,m)$ is nominally its weak-spin-orbit local electron orbital  quantum number set [or its fully relativistic set $(n,j)$], $\psi^{\dag}_{j,\sigma}({\bm r}_j)$ creates an electron or hole with spin eigenstate $\sigma=\pm1$ at position ${\bm r}_j$ in the $j^{\rm th}$ conduction band $\omega_{\rm n}={\bm \mu}_{\rm n}\cdot{\bm B}_0$, $\omega_{\rm e}={\bm \mu}_{\rm e}\cdot{\bm B}_0$, $\omega'_{j,{\rm e}}={\bm \mu}_e\cdot{\bm g}_j\cdot{\bm B}_0$ are respectively the Zeeman energies for the probed nucleus, local orbital electrons, and conduction electrons, respectively, where ${\bm \mu}_{\rm n}$ is the nuclear magneton for the probed nucleus (the magnitude of which can be positive or negative), $|{\bm \mu}_{\rm e}|=\mu_B$ is the Bohr magneton, ${\bm g}_j\cdot{\bm B}_0$ defines the quantization axis direction for the anisotropic but assumed  diagonal ${\bm g}_j$ tensor in the $j^{\rm th}$ of the $N_b$ conduction bands with effective mass $m_{j,\nu}$ in the $\nu^{\rm th}$ spatial direction,  ${\bm A}_{j,\nu}({\rm r}_j)$ is the magnetic vector potential at the position ${\bm r}_j$ of the conduction electron in the $\nu^{\rm th}$ band, the time independent magnetic induction ${\bm B}_0={\bm \nabla}_{j,\nu}\times{\bm A}_{j,\nu}$ is the same in each band,  ${\rm i}=\sqrt{-1}$,  and we set $\hbar=1$. Here we use the previous notation\cite{Hall2016}, but rearrange the terms in the overall Hamiltonian in order to properly take account of both the time $t$ and temperature $T$ dependencies essential for probing the superconducting state.  We note that for integer or half-integer $I$, the nuclei would normally be expected to obey Bose-Einstein or Fermi-Dirac statistics, but since different nuclei correspond to different atoms and do not come in contact with one another, that statistics is not expected to be an important feature of the Knight shift.  Equation 1 is the extension to arbitrary nuclear spin $I$ of the bare Hamiltonian studied previously, except that ${\cal H}_{{\rm cond},0}$ was the time independent part of ${H}_{A,2}$\cite{Hall2016}.  We note that for a diagonal ${\bm g}_j$ tensor,
\begin{eqnarray}
\omega'_{j,{\rm e}}&=&\mu_BB_0[g_{j,xx}\sin^2\theta\cos^2\phi+g_{j,yy}\sin^2\theta\sin^2\phi+g_{j,zz}\cos^2\theta]^{1/2}.
\end{eqnarray}

As a starting point, we assume ${\bm B}_0$ is uniform in the probed material, but when the material goes into the superconducting state, and ${\bm B}_0$ is in an arbitrary direction with respect to the crystal axes, this is only true at the upper critical field ${\bm H}_{c2}$ above which the superconductor becomes a normal metal\cite{Klemm2012,Klemm1980}.  However, in the mixed state for which the time-independent part of the applied magnetic field ${\bm H}_0$ satisfies ${\bm H}_{c1}<{\bm H}_0<{\bm H}_{c2}$, if ${\bm H}_0$ is along a crystal axis, the direction of ${\bm B}_0$ is the same as the direction of ${\bm H}_0$ \cite{Klemm2012,Klemm1993}.

\subsection{The Time-Independent Interaction Hamiltonian ${\cal H}_{\rm int}$}
We write the  time-independent interaction part ${\cal H}_i$ of the Hamiltonian as
\begin{eqnarray}
{\cal H}_{\rm int}&=&{\cal H}_{hf}+{\cal H}_{{\rm e, int}}+{\cal H}_{\rm e, cond}+{\cal H}_{{\rm sc}}\hskip10pt,\label{eqn5}
\end{eqnarray}
where
\begin{eqnarray}
{\cal H}_{hf}&=&-\frac{D_z}{4}\sum_{i,q,\sigma,m_I}m_I\sigma a^{\dag}_{i,m_I}a^{}_{i,m_I}b^{\dag}_{i,q,\sigma}b^{}_{i,q,\sigma}-\frac{D_x}{2}\sum_{i,q,\sigma,m_I}A^{\sigma}_{I,m_I}a^{\dag}_{i,m_I+\sigma}a^{}_{i,m_I}b^{\dag}_{i,q,-\sigma}b^{}_{i,q,\sigma}\hskip10pt,\\
{\cal H}_{\rm e,int}&=&\frac{1}{2}\sum_{i,q,\sigma}U_q\hat{n}_{i,q,\sigma}\hat{n}_{i,q,-\sigma}\hskip10pt =\sum_{i,q}U_qb^{\dag}_{i,q,\uparrow}b^{}_{i,q,\uparrow}b^{\dag}_{i,q,\downarrow}b^{}_{i,q,\downarrow},\\
{\cal H}_{\rm e,cond}&=&\sum_{i,q,j,\sigma}\int d^3{\bm r}_j\Bigl(\nu_{i,q,j}\psi^{\dag}_{j,\sigma}({\bm r}_j)b_{i,q,\sigma}+H.c.\Bigr)\delta^{(3)}({\bm r}_j-{\bm r}_i)\hskip10pt,
\end{eqnarray}
where $A^{\sigma}_{I,m_I}=\sqrt{I(I+1)-m_I(m_I+\sigma)}$,
and depending upon what is calculated, the superconducting pairing interaction may be written either in real space as
\begin{eqnarray}
{\cal H}^{\rm pos}_{\rm sc}&=&\frac{1}{2}\sum_{j,j',\sigma,\sigma'}\int d^3{\bm r}_j\int d^3{\bm r}'_{j'}\psi^{\dag}_{j,\sigma}({\bm r}_j)\psi^{\dag}_{j',\sigma'}({\bm r}'_{j'})V_{j,j';\sigma,\sigma'}({\bm r}_j-{\bm r}'_{j'})\psi^{}_{j',\sigma'}({\bm r'}_{j'})\psi^{}_{j,\sigma}({\bm r}_j)\hskip2pt,
\end{eqnarray}
or in momentum space as
\begin{eqnarray}
{\cal H}^{\rm mom}_{\rm sc}&=&\frac{1}{2}\sum_{j,j',\sigma,\sigma'}\int\frac{d^3{\bm k}_j}{(2\pi)^3}\int\frac{d^3{\bm k}'_{j'}}{(2\pi)^3}\psi^{\dag}_{j,\sigma}({\bm k}_j)\psi^{\dag}_{j',\sigma'}({\bm k}'_{j'})V_{j,j';\sigma,\sigma'}({\bm k}_j-{\bm k}'_{j'})\psi^{}_{j',\sigma'}({\bm k}'_{j'})\psi^{}_{j,\sigma}({\bm k}_j)\hskip2pt.
\end{eqnarray}
Although it appears at first sight to be easier to extend the calculation of the Knight shift into the BCS superconducting state by using ${\cal H}^{\rm pos}_{\rm sc}$ in order to include the Zeeman terms, we have included the momentum-space pairing interaction ${\cal H}^{\rm mom}_{\rm sc}$ for  $p$-wave superconductors in magnetic fields \cite{Scharnberg1980,Scharnberg1985}, for which the simplest single-band parallel-spin pairing interaction $V_{j,j';\sigma,\sigma'}({\bm k}_j-{\bm k}'_{j'})=-V_0\delta_{j,j'}\delta_{j,1}\delta_{\sigma,\sigma'}{\bm k}_1\cdot{\bm k}'_1$ \cite{Scharnberg1980}, and a modification of  ${\cal H}^{\rm mom}_{\rm sc}$ more naturally treats the pairing of conduction electrons (or holes) in the presence of a strong ${\bm B}_0$.

We note that $\sigma=\pm1$ present in $A^{\sigma}_{I,m_I}$ corresponds to the correct matrix elements for raising and lowering the $m_I$ value and also corresponds to our description of the  spin-1/2 electron spins\cite{Hall2016}. Of course, $m_I$ and $\sigma$ are restricted by $-I\le m_I,m_I+\sigma\le I$.  The first three of these terms were presented previously\cite{Hall2016}, except for a slightly different normalization factor proportional to $N_b$, and respectively represent the hyperfine interaction between the nuclear and surrounding orbital electrons, the effective local electron correlation interaction, and the effective Anderson interaction that allows an orbital electron to leave a local atomic site and jump into a conduction band \cite{Anderson1961}.  The last term ${\cal H}_{\rm sc}$ is responsible for superconducting pairing, and in the form presented allows for pairing between electrons or holes in different bands and with either the same ($\sigma'=\sigma$) or different ($\sigma'=-\sigma$) spins.  In most superconductors, interband pairing is generally considered to be less important than is intraband pairing, but such complications might be important in cases such as Sr$_2$RuO$_4$, for which two of the bands are nearly identical. For standard BCS pairing, we would have

$V_{\sigma,\sigma'}({\bm r}_j-{\bm r}'_{j'})\rightarrow-V_0\delta_{\sigma',-\sigma}\delta_{j,j'}\delta^{(3)}({\bm r}_j-{\bm r}'_{j}),$
at least in the standard approximation.  For parallel-spin $p$-wave superconductors, one cannot assume the paired electrons are at the same location,  but  different approximations have been found to give reliable results for the upper critical induction in ferromagnetic superconductors\cite{Scharnberg1980,Scharnberg1985,Zhang2014a,Loerscher2013, Zhang2014b}.

\subsection{The Time-Dependent Hamiltonian ${\cal H}'(t)$}
The crucial part of a magnetic resonance experiment arises from the time-dependent field transverse to the stronger constant magnetic field.   In a conventional NMR experiment, the time-dependent induction ${\bm B}_1(t)$ oscillates in the plane normal to the strong, constant magnetic induction ${\bm B}_0$.  For ${\bm B}_0=B_0(\sin\theta\cos\phi,\sin\theta\sin\phi,\cos\theta)=B_0\hat{\bm r}$,   one may then write ${\bm B}_1(t)=B_1\{\cos[\omega_0(t-t_0)]\hat{\bm \theta}-\sin[\omega_0(t-t_0)]\hat{\bm\phi}\}$, where $\hat{\bm \theta}=(\cos\theta\cos\phi,\cos\theta\sin\phi,-\sin\theta)$ and $\hat{\bm \phi}=(-\sin\phi,\cos\phi,0)$ in the same Cartesian coordinates, and in order not to get confused with the time contours, we may choose ${\bm B}_1(t_0)$ to be along $\hat{\bm \theta}$.  This is the classic way to obtain a resonance in the power spectrum associated with flipping an electron or proton spin from up to down, or in a spin $I$ nucleus, to obtain a regular pattern of resonance frequencies associated with changes in the multiple Zeeman-like nuclear spin levels.  Since one generally takes $B_1\ll B_0$, this classic case is generally adiabatic\cite{Berry1984,Griffiths2005}, and is the simplest case to treat analytically.

For the above classic NMR case of a single angular frequency $\omega_0$ in ${\bm B}_1(t)$, we then have
\begin{eqnarray}
{\cal H}'(t)&=&{\cal H}_{\rm n}'(t)+{\cal H}'_{\rm e}(t)+{\cal H}'_{\rm cond}(t)\hskip10pt,\\
{\cal H}'_{\rm n}(t)&=&-\frac{\Omega_{\rm n}}{2}\sum_{i,m_I,\sigma}e^{{\rm i}\sigma\omega_0(t-t_0)}A^{\sigma}_{I,m_I}a^{\dag}_{i,m_I+\sigma}a^{}_{i,m_I}\hskip10pt,\\
{\cal H}'_{\rm e}(t)&=&-\frac{\Omega_{\rm e}}{2}\sum_{i,\sigma}e^{{\rm i}\sigma\omega_0(t-t_0)}b^{\dag}_{i,\sigma}b^{}_{i,-\sigma}\hskip10pt,\\
{\cal H}'_{\rm cond}(t)&=&-\frac{1}{2}\sum_{j,\sigma}e^{{\rm i}\sigma\omega_0(t-t_0)}\int d^3{\bm r}_j\psi^{\dag}_{j,\sigma}({\bm r}_j)\Omega'_{j,{\rm e}}\psi^{}_{j,-\sigma}({\rm r}_j)\hskip10pt,
\end{eqnarray}
where $\Omega_{\rm n}=\mu_{\rm n}B_1$,   $\Omega_{\rm e}=\mu_{\rm e}B_1$, and $\Omega'_{j,{\rm e}}={\bm\mu}_e\cdot{\bm g}_j\cdot{\bm B}_1$, and $A^{\sigma}_{I,m_I}$ is given by Equation 11.\cite{Hall2016}.

\section{The Keldysh contours}
Following Haug and Jauho \cite{Haug2008} and with regard to the contours, Rammer and Smith \cite{Rammer1986}, we may treat the time and temperature dependence of the particles together in the same formulas, as long as we properly order the time integrations around the appropriate contours.  When there is only one type of particle, which we take to be a fermion, the fields at the three-dimensional positions ${\bm r}_1$ and ${\bm r}_{1'}$ evolve in time according to the simple Hamiltonian $H_0$,
\begin{eqnarray}
\psi^{}_{{\cal H}_0}({\bm r}_1,t_1)&\equiv&\psi^{}_{{\cal H}_0}(1)=e^{{\rm i}{\cal H}_0t_1}\psi^{}({\bm r}_1)e^{-{\rm i}{\cal H}_0t_1},\\
\psi^{\dag}_{{\cal H}_0}({\bm r}_{1'},t_{1'})&\equiv&\psi_{{\cal H}_0}^{\dag}(1')=e^{{\rm i}{\cal H}_0t_{1'}}\psi^{\dag}({\bm r}_{1'})e^{-{\rm i}{\cal H}_0t_{1'}},
\end{eqnarray}
with its density matrix also involving only  ${\cal H}_0$ (and the number operator $\hat{\cal N}$ in the grand canonical ensemble statistics),
\begin{eqnarray}
\hat{\rho}_0&=&\frac{e^{-\beta({\cal H}_0-\mu\hat{\cal N})}}{{\rm Tr}[e^{-\beta({\cal H}_0-\mu\hat{\cal N})}]},
\end{eqnarray}
and the Green function is given by the two contour integration paths $C$ and $C_{\rm int}$,
\begin{eqnarray}
G(1,1')&=&-{\rm i}\frac{{\rm Tr}\Bigl\{\hat{\rho}_0T_{C}\bigl[{\cal S}_{C_{\rm int}}{\cal S}'_C\psi^{}_{{\cal H}_0}(1)\psi^{\dag}_{{\cal H}_0}(1')\bigr]\Bigr\}}{{\rm Tr}\bigl[T_C({\cal S}_{C_{\rm int}}{\cal S}'_C)\bigr]},\\
{\cal S}'_C&=&\exp\bigl[-{\rm i}\int_Cd\tau{\cal H}'_{{\cal H}_0}(\tau)\bigr],\\
{\cal S}_{{C_{int}}}&=&\exp\bigl[-{\rm i}\int_{C_{\rm int}}d\tau{\cal H}_{{\rm int},{\cal H}_0}(\tau)\bigr],
\end{eqnarray}
where the operators in ${\cal H}_{{\cal H}_0}'(\tau)$ and ${\cal H}_{{\rm int},{\cal H}_0}(\tau)$ evolve in time via the easily soluble Hamiltonian ${\cal H}_0$.  The Greek letter $\tau$ implies that one needs to consider it as being just above or just below the real axis until the contours merge into one.  Roman lettering ($t$) indicates the integrals are on the real axis.

For the case of the time-dependent Hamiltonian ${\cal H}'(t)$ making adiabatic changes, as in the case considered here, the two contours $C_{\rm int}$ and $C$ shown in Figure 1 merge into the left contour $C$.  We use the standard short-hand notation
\begin{eqnarray}
G(1,1')&=&-{\rm i}\langle T_C[\psi_{\cal H}(1)\psi^{\dag}_{\cal H}(1')]\rangle,
\end{eqnarray}
where the particle type, its position, and its energy are still undefined.  In order to treat the various time orderings on the contour $C$, we define in the Heisenberg representation for the full Hamiltonian ${\cal H}$,
\begin{eqnarray}
G^{>}(1,1')&=&-{\rm i}\langle\psi_{\cal H}(1)\psi^{\dag}_{\cal H}(1')\rangle,\\
G^{<}(1,1')&=&+{\rm i}\langle\psi^{\dag}_{\cal H}(1')\psi_{\cal H}(1)\rangle,\\
G^{C}(1,1')&=&-{\rm i}\langle T[\psi_{\cal H}(1)\psi^{\dag}_{\cal H}(1')]\rangle=\Theta(t_1-t_{1'})G^{>}(1,1')+\Theta(t_{1'}-t_1)G^{<}(1,1'),\\
G^{\tilde{C}}(1,1')&=&-{\rm i}\langle \tilde{T}[\psi_{\cal H}(1)\psi^{\dag}_{\cal H}(1')]\rangle=\Theta(t_1-t_{1'})G^{<}(1,1')+\Theta(t_{1'}-t_1)G^{>}(1,1'),
\end{eqnarray}
where the ordinary time-ordering operator $T$ and inverse-time-ordering operator $\tilde{T}$ describe opposite directions in time, as sketched by lines $C_1$ and $C_2$ in Figure 2.  We note that $G^C(1,1')+G^{\tilde{C}}(1,1)=G^{<}(1,1')+G^{>}(1,1')$, so only three of these Green functions are linearly independent.

Here we need to describe three particles, all of which are effectively fermions.
\begin{figure}
\center{\includegraphics[width=0.4\textwidth]{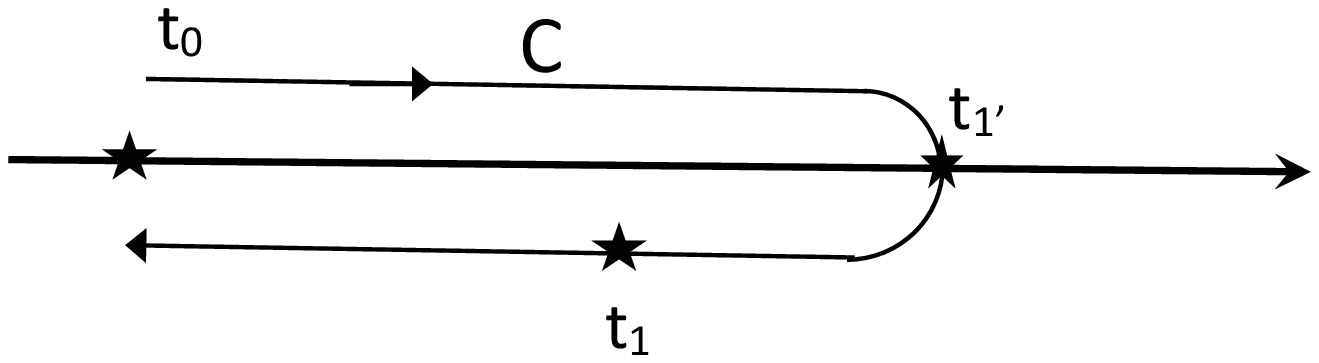}\hskip10pt\includegraphics[width=0.4\textwidth]{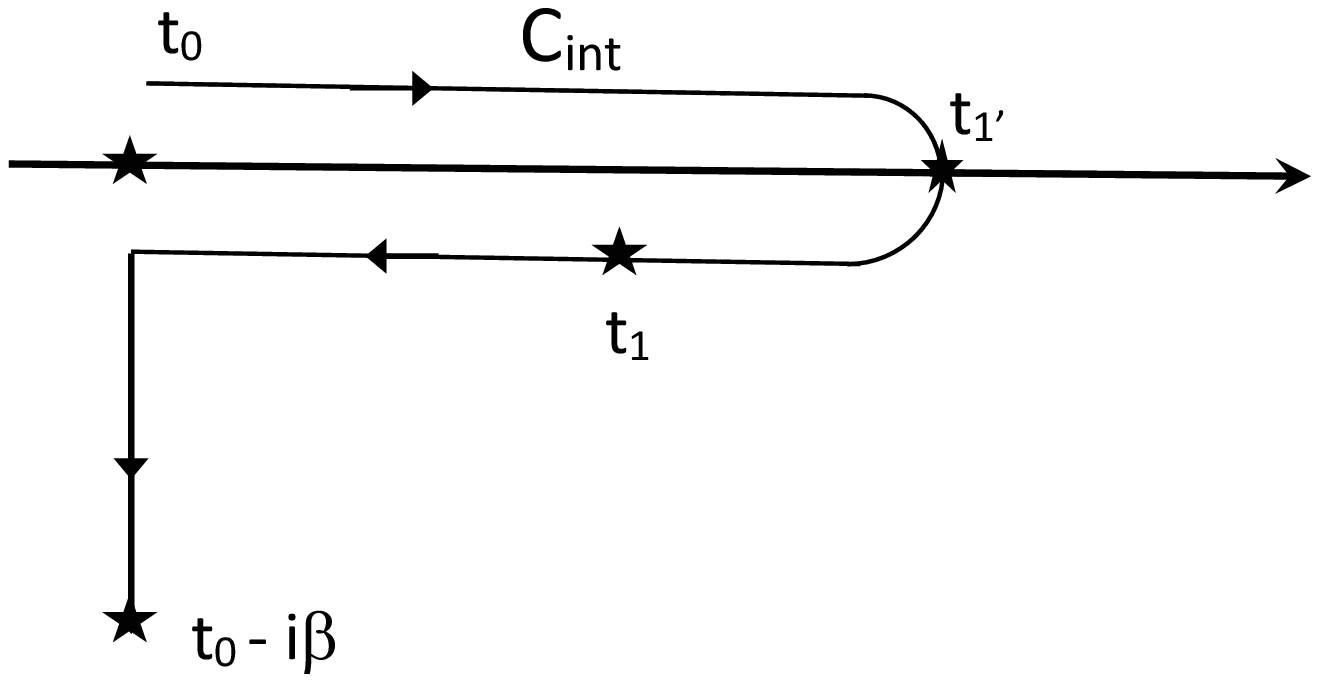}
\caption{Left: Sketch of the ``closed path'' contour C. Right: Sketch of the ``interaction'' Contour C$_{\rm int}$.}}\label{fig1}
\end{figure}
\begin{figure}
\center{\includegraphics[width=0.4\textwidth]{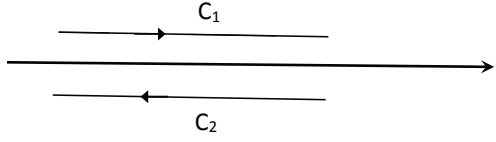}
\caption{Sketch of the Keldysh  contour C$_K$.}}\label{fig2}
\end{figure}

\subsection{Bare Nuclear Contour Green Functions}
We first consider the nuclei, which are assumed not to interact with one another, as they are fixed in the crystalline locations, which if there is more than one isotope of a particular type with spin $I$, may be at a random selection  of crystalline sites.  In the presence of the constant magnetic induction ${\bm B}_0$, it can be in any one of the $2I+1$ manifold of nuclear Zeeman states, but because each of these local states at the probed nuclear site $i$ can be either  unoccupied or singly occupied, this manifold of local nuclear spin states is precisely that of a fermion with $(2I+1)$ states.  Its occupancy in the local state $m_I$ on site $i$ in the grand canonical ensemble is therefore easily seen to be
\begin{eqnarray}
\langle\hat{n}^{\rm n}_{i,m_I}\rangle&=&\frac{1}{e^{\beta(\epsilon_{m_I}-\mu_{\rm ncp})}+1},
\end{eqnarray}
where $\epsilon_{m_I}=-\omega_{\rm n}m_I$ and $\mu_{\rm ncp}$ is the nuclear chemical potential.  We then have for the bare nuclear Green functions with ${\cal H}={\cal H}_{0,{\rm n}}$,
\begin{eqnarray}
G^{(0,{\rm n}),<}_{i,i';m_I,m_I'}(1,1')&=&+{\rm i}\delta_{i,i'}\delta_{m_I,m_I'}e^{{\rm i}\epsilon_{m_I}(t_1-t_{1'})}\langle\hat{n}^{\rm n}_{i,m_I}\rangle,\\
G^{(0,{\rm n}),>}_{i,i';m_I,m_I'}(1,1')&=&-{\rm i}\delta_{i,i'}\delta_{m_I,m_I'}e^{{\rm i}\epsilon_{m_I}(t_1-t_{1'})}[1-\langle\hat{n}^{\rm n}_{i,m_I}\rangle],
\end{eqnarray}
and $G^{(0,{\rm n}),C}_{i,i';m_I,m_I'}(1,1'), G^{(0,{\rm n}),\tilde{C}}_{i,i';m_I,m_I'}(1,1')$ are constructed  from these according to Equations 25 and 26.
There are only three distinct bare neutron Green functions.  This is also true when interactions are included \cite{Haug2008}.  Although it is somewhat surprising that the nuclear occupation density has the Fermi function form even for integral spin $I$, this is due to the nuclear Zeeman magnetic level occupancy  being either 0 or 1 for each level on a given probed nuclear site.
\subsection{Bare Orbital Electron Contour Green Functions}
 For the surrounding orbital electrons, we assume that the  magnetic induction ${\bm B}_0+{\bm B}_1(t)$ is sufficiently weak that it does not change the electronic structure of the orbital electrons or lead to transitions between the orbital electron states and energy levels.  Thus, we assume that it only interacts with the orbital electron spins.  We note that this is expected to be a good approximation, as the total charge of the nucleus plus its orbital electrons is on the order of one electron charge (for an ion), and the mass of the ion is so large that any Landau levels describing the orbital electrons and their central nucleus is completely negligible in comparison with the Landau levels of the conduction electrons.  The only point then to consider for the interaction of ${\bm B}_0$ with the orbital electrons is that there can be either 0, 1, or 2 electrons in a given orbital energy $\epsilon_q$, and two possible magnetic energies for up and down spins.  Hence, it is elementary to show that average orbital electron occupation number in the grand canonical ensemble is
\begin{eqnarray}
\langle\hat{n}^{\rm e}_{i,q,\sigma}\rangle&=&\frac{1}{e^{\beta(\epsilon_{q}-\sigma\omega_{\rm e}/2-\mu_{\rm ecp})}+1},
\end{eqnarray}
where $\mu_{\rm ecp}$ is the orbital electron chemical potential.  It is then easy to show that the bare orbital electron Green functions are
\begin{eqnarray}
G^{(0,{\rm e}),<}_{i,i';q,q'\atop{\sigma,\sigma'}}(1,1')&=&+{\rm i}\delta_{i,i'}\delta_{q,q'}\delta_{\sigma,\sigma'}e^{{\rm i}(\epsilon_{q}-\sigma\omega_{\rm e}/2)(t_1-t_{1'})}\langle\hat{n}^{\rm e}_{i,q,\sigma}\rangle,\\
G^{(0,{\rm e}),>}_{i,i';q,q'\atop{\sigma,\sigma'}}(1,1')&=&-{\rm i}\delta_{i,i'}\delta_{q,q'}\delta_{\sigma,\sigma'}e^{{\rm i}(\epsilon_{q}-\sigma\omega_{\rm e}/2)(t_1-t_{1'})}[1-\langle\hat{n}^{\rm e}_{i,q,\sigma}\rangle],
\end{eqnarray}
and the time-ordered and inverse-time-ordered bare orbital electron Green functions are obtained analogously to Equations 25 and 26.  Only three the bare orbital electron Green functions are linearly independent.  This is also true when interactions are included \cite{Haug2008}.
\subsection{Bare Conduction Electron  Contour Green Functions}
In a normal metal (all superconductors including the cuprates and the record high transition temperature superconductor hydrogen sulfide, which probably transforms to H$_3$S under the 155 GPa pressure that causes it to become superconducting at 203 K\cite{Drozdov2015}), the conduction electrons or holes propagate throughout the metal with wave vectors on or nearly on one or more Fermi surfaces.  Both the spins and the  charges of the conduction electrons interact with ${\bm B}_0$, the spins via the Zeeman interaction and the charges couple to the magnetic vector potential, leading to  Landau orbits.   Here we assume each of these potentially multiple Fermi surfaces has an ellipsoidal shape, but the shapes and orientations of each of the Fermi surfaces can be different from one another.

We first use the Klemm-Clem transformations to transform each of the ellipsoidal conduction electron band dispersions into spherical forms \cite{Klemm1980,Hall2016}.  For each ellipsoidal band, the anisotropic scale transformation that preserves the Maxwell equation ${\bm\nabla}\cdot{\bm B}_0=0$ transforms the elliptical Fermi surface into a spherical one, but rotates the transformed induction differently in each band.  Then, one rotates these bands so that the rotated induction is along the $z$ direction in each band \cite{Hall2016}. In the $j^{\rm th}$ band, the conduction electrons behave as free particles with wave vector $k_{j,||}$ along the transformed $\hat{z}$ direction, but propagate in Landau orbits indexed by the harmonic oscillator quantum number $n_j$.  Thus, we need to requantize the conduction electron fields as $\tilde{\psi}_{j,n_j,\sigma}(k_{j,||})$.

We therefore rewrite the transformed $\tilde{\cal H}_{{\rm cond},0}$ as
\begin{eqnarray}
\tilde{\cal H}_{\rm cond,0}&=&\sum_{j,\sigma}g_{{\rm L},j}\sum_{n_j=0}^{\infty}\int\frac{dk_{j,||}}{2\pi}\tilde{\psi}^{\dag}_{j,n_j,\sigma}(k_{j,||})[\varepsilon_j(n_j,k_{j,||})-\sigma\tilde{\omega}'_{j,{\rm e}}/2]\tilde{\psi}^{}_{j,n_j,\sigma}(k_{j,||}),\\
\varepsilon_j(n_j,\tilde{k}_{j,||})&=&\frac{\tilde{k}_{j,||}^2}{2m_j\alpha_j^2}+\frac{(n_j+1/2){\rm e}B_0\alpha_j}{m_j},\\
\tilde{\omega}'_{j,{\rm e}}&=&\mu_BB_0\beta_j(\theta,\phi),
\end{eqnarray}
where $n_j=0,1,2,\ldots$ are the two-dimensional simple harmonic oscillator quantum numbers of the Landau orbits for band $j$, $k_{j,||}$ are the free-particle dispersions along the transformed induction direction,
\begin{eqnarray}
\alpha_j(\theta,\phi)&=&[\overline{m}_{j,1}\sin^2\theta\cos^2\phi+\overline{m}_{j,2}\sin^2\theta\sin^2\phi+\overline{m}_{j,3}\cos^2\theta]^{1/2},\\
m_j&=&(m_{j,1}m_{j,2}m_{j,3})^{1/3},\\
\overline{m}_{j,\nu}=\frac{m_{j,\nu}}{m_j},\\
\beta_j(\theta,\phi)&=&[g^2_{j,xx}\overline{m}_{j,1}\sin^2\theta\cos^2\phi+g^2_{j,yy}\overline{m}_{j,2}\sin^2\theta\sin^2\phi+g^2_{j,zz}\overline{m}_{j,3}\cos^2\theta]^{1/2},
\end{eqnarray}
and
\begin{eqnarray}
g_{{\rm L},j}&=&\frac{eB_0\alpha_j}{2\pi}
\end{eqnarray}
is the spatially-transformed Landau degeneracy  for a single electron in the $j^{\rm th}$ band.
We may then write the conduction electron occupation number as
\begin{eqnarray}
\langle\hat{n}^{\rm cond}_{j,\tilde{k}_{j,||},n_j,\sigma}\rangle&=&\frac{1}{e^{\beta[\varepsilon_j(n_j,\tilde{k}_{j,||})-\sigma\tilde{\omega}'_{j,{\rm e}}/2-\mu_{\rm cond, cp}]}+1}
\end{eqnarray}
for a diagonal ${\bm g}_j$ tensor describing the spins of the $j^{\rm th}$ conduction band, where $\mu_{\rm cond,cp}$ is the chemical potential of the conduction electrons.  We note that all of the bands that cross this conduction electron chemical potential make important contributions to the Knight shift.

The bare conduction electron Green functions can then be found

\begin{eqnarray}
G^{(0,{\rm cond}),<}_{j,j';\tilde{k}_{j,||},\tilde{k}'_{j',||};\atop{n_j,n'_{j'};\sigma,\sigma'}}(1,1')&=&+{\rm i}\delta_{j,j'}\delta_{\tilde{k}_{j,||},\tilde{k}'_{j',||}}\delta_{n_j,n'_{j'}}\delta_{\sigma,\sigma'}g_{{\rm L},j}e^{{\rm i}[\varepsilon_j(\tilde{k}_{j,||},n_j)-\sigma\tilde{\omega}'_{\rm e}/2](t_1-t_{1'})}\langle\hat{n}^{\rm cond}_{i,\tilde{k}_{j,||},n_j,\sigma}\rangle,\\
G^{(0,{\rm cond}),>}_{j,j';\tilde{k}_{j,||},\tilde{k}'_{j',||};\atop{n_j,n'_{j'};\sigma,\sigma'}}(1,1')&=&-{\rm i}\delta_{j,j'}\delta_{\tilde{k}_{j,||},\tilde{k}'_{j',||}}\delta_{n_j,n'_{j'}}\delta_{\sigma,\sigma'}g_{{\rm L},j}e^{{\rm i}[\varepsilon_j(\tilde{k}_{j,||},n_j)-\sigma\tilde{\omega}'_{\rm e}/2](t_1-t_{1'})}[1-\langle\hat{n}^{\rm cond}_{i,\tilde{k}_{j,||},n_j,\sigma}\rangle],
\end{eqnarray}
and the contour-ordered and inverse-contour-ordered bare conduction electron Green functions are obtained as in Equations 25 and 26, so that there are only three independent bare  conduction electron Green functions.

 Furthermore, due to the strong ${\bm B}_0$, we also need to spatially transform all of the other terms in the Hamiltonian that contain the conduction electrons.  Thus, we have \cite{Hall2016}
\begin{eqnarray}
\tilde{\cal H}'_{\rm cond}(t)&=&-\frac{1}{2}\sum_{j,\sigma}g_{{\rm L}, j}e^{{\rm i}\sigma\omega_0(t-t_0)}\sum_{n_j=0}^{\infty}\int\frac{d k_{j,||}}{2\pi}\tilde{\psi}^{\dag}_{j,n_j,\sigma}(k_{j,||})\tilde{\Omega}'_{j,{\rm e}}\tilde{\psi}^{}_{j,n_j,-\sigma}(k_{j,||}),\\
\tilde{\Omega}'_{j,{\rm e}}&\approx&\mu_BB_1\gamma_j(\theta,\phi),\\
\gamma_j(\theta,\phi)&=&[g^2_{j,xx}(\overline{m}_{j,2}\sin^2\theta\sin^2\phi+\overline{m}_{j,3}\cos^2\theta)\nonumber\\
&&+g^2_{j,yy}(\overline{m}_{j,1}\sin^2\theta\cos^2\phi+\overline{m}_{j,3}\cos^2\theta)+g^2_{j,zz}(\overline{m}_{j,1}\cos^2\phi+\overline{m}_{j,2}\sin^2\phi)]^{1/2}.
\end{eqnarray}

 \section{Transformations in Time of the Operators with the Bare Hamiltonian}
 In order to proceed with the perturbation expansions, we first need to transform the nuclear, orbital electronic, and conduction electronic operators in real time, using the bare Hamiltonian in Equation 1.  For the nuclear and orbital electronic operators, this is elementary.  We have
 \begin{eqnarray}
 a^{}_{i,m_I}(t)&=&e^{{\rm i}{\cal H}_{{\rm n},0}t}a^{}_{i,m_I}e^{-{\rm i}{\cal H}_{{\rm n},0}t}\nonumber\\
 \frac{d a_{i,m_I}(t)}{dt}&=&{\rm i}\left[{\cal H}_{n,0},a^{}_{i,m_I}(t)\right]\nonumber\\
 &=&+{\rm i}\omega_{\rm n}m_Ia^{}_{i,m_I}(t),
 \end{eqnarray}
 and integrating the elementary differential equation, we immediately find
 \begin{eqnarray}
 a^{}_{i,m_I}(t)&=&e^{{\rm i}\omega_{\rm n}m_It}a^{}_{i,m_I}(0)=e^{{\rm i}\omega_{\rm n}m_I(t-t_0)}a^{}_{i,m_I}(t_0),
 \end{eqnarray}
 in order to use this in Equation 16.  The quantity $a^{\dag}_{i,m_I+\sigma}(t)$ is instantly obtained from the Hermitian conjugate of Equation 52 and letting $m_I\rightarrow m_I+\sigma$, and hence $a^{\dag}_{i,m_I+\sigma}(t)=e^{-{\rm i}\omega_{\rm n}(m_I+\sigma)(t-t_0)}a^{\dag}_{i,m_I+\sigma}(t_0)$, so that the time-transformed Equation 16 becomes
 \begin{eqnarray}
 {\cal H}'_{\rm n,{\cal H}_{{\rm n},0}}(t)&=&-\frac{\Omega_{\rm n}}{2}\sum_{i,m_I,\sigma}e^{{\rm i}\sigma(\omega_0-\omega_{\rm n})(t-t_0)}A^{\sigma}_{I,m_I}a^{\dag}_{i,m_I,\sigma}(t_0)a_{i,m_I}(t_0).
 \end{eqnarray}
 Similarly, for the local orbital electron operators, we have
 \begin{eqnarray}
 b_{i,q,\sigma}(t)&=&e^{{\rm i}{\cal H}_{{\rm e},0}(t-t_0)}b_{i,q,\sigma}(t_0)e^{-{\rm i}{\cal H}_{{\rm e},0}(t-t_0)}\\
 &=&e^{-{\rm i}(\varepsilon_q-\sigma\omega_{\rm e}/2)(t-t_0)}b_{i,q,\sigma}(t_0),
 \end{eqnarray}
 and
 \begin{eqnarray}
 {\cal H}'_{{\rm e},{\cal H}_{{\rm e},0}}(t)&=&-\frac{\Omega_{\rm e}}{2}\sum_{i,q,\sigma}e^{{\rm i}\sigma(\omega_0-\omega_{\rm e})(t-t_0)}b^{\dag}_{i,q,\sigma}(t_0)b_{i,q,-\sigma}(t_0).
 \end{eqnarray}
 For the  spatially-transformed conduction electron operators,
 \begin{eqnarray}
 \tilde{\psi}^{}_{j,n_j,\sigma}(k_{j,||},t)&=&e^{{\rm i}\tilde{\cal H}_{\rm cond,0}(t-t_0)}\tilde{\psi}_{j,n_j,\sigma}(k_{j,||},t_0)e^{-{\rm i}\tilde{\cal H}_{\rm cond,0}(t-t_0)}\\
 &=&e^{-{\rm i}[\varepsilon_j(n_j,k_{j,||})-\sigma\tilde{\omega}'_{j,{\rm e}}/2](t-t_0)}\tilde{\psi}^{}_{j,n_j,\sigma}(k_{j,||},t_0),
 \end{eqnarray}
 so that the time-transformed Equation 44 becomes
 \begin{eqnarray}
 \tilde{\cal H}'_{{\rm cond},\tilde{\cal H}_{\rm cond,0}}(t)&=&-\frac{1}{2}\sum_{j,\sigma}g_{{\rm L},j}\sum_{n_j=0}^{\infty}\int\frac{dk_{j,||}}{2\pi}e^{{\rm i}\sigma(\omega_0-\tilde{\omega}'_{j,{\rm e}})(t-t_0)}\tilde{\psi}^{\dag}_{j,n_j,\sigma}(k_{j,||},t_0)\tilde{\Omega}'_{j,{\rm e}}\tilde{\psi}^{}_{j,n_j,-\sigma}(k_{j,||},t_0).\nonumber\\
 \end{eqnarray}
 We note that all three of these transformed Hamiltonians correspond to spin-dependent external field interactions, where the fields are
 \begin{eqnarray}
 U_{{\rm n},i,i';m_I,m_I'}(t)&=&-\frac{\Omega_{\rm n}}{2}\delta_{i,i'}\sum_{\sigma=\pm1}\delta_{m_I',m_I+\sigma}A^{\sigma}_{I,m_I}e^{{\rm i}\sigma(\omega_0-\omega_{\rm n})(t-t_0)},\\
 U_{{\rm e},i,i';q,q'\atop{\sigma,\sigma'}}(t)&=&-\frac{\Omega_{\rm e}}{2}\delta_{i,i'}\delta_{q,q'}\delta_{\sigma',-\sigma}e^{{\rm i}\sigma(\omega_0-\omega_{\rm e})(t-t_0)},\\
 U_{{\rm cond}, j,j'; n_j,n_j'\atop{ k_{j,||},k'_{j,||}};\sigma,\sigma'}(t)&=&-\frac{\tilde{\Omega}'_{j,{\rm e}}}{2}\delta_{j,j'}\delta_{n_j,n'_j}\delta_{k_{j,||},k'_{j,||}}\delta_{\sigma',-\sigma}e^{{\rm i}\sigma(\omega_0-\tilde{\omega}'_{j,{\rm e}})(t-t_0)}.
 \end{eqnarray}
 Then, we time transform the difficult (interaction) parts of the full Hamiltonian.  The hyperfine and local electron-electron interactions are elementary to transform.  We obtain
 \begin{eqnarray}
 {\cal H}_{hf,{\cal H}_{{\rm n},0}+{\cal H}_{{\rm e},0}}(t)&=&-\frac{D_z}{4}\sum_{i,q,\sigma,m_I}m_I\sigma a^{\dag}_{i,m_I}(t_0)a^{}_{i,m_I}(t_0)b^{\dag}_{i,q,\sigma}(t_0)b^{}_{i,q,\sigma}(t_0)\nonumber\\
 & &-\frac{D_x}{2}\sum_{i,q,\sigma,m_I}A^{\sigma}_{I,m_I}e^{{\rm i}\sigma(\omega_{\rm e}-\omega_{\rm n})(t-t_0)}a^{\dag}_{i,m_I+\sigma}(t_0)a^{}_{i,m_I}(t_0)b^{\dag}_{i,q,-\sigma}(t_0)b^{}_{i,q,\sigma}(t_0),\\
 {\cal H}_{{\rm e, int},{\cal H}_{{\rm e,0}}}(t)&=&\frac{1}{2}\sum_{i,q,\sigma}U_q\hat{n}_{i,q,\sigma}(t_0)\hat{n}_{i,q,-\sigma}(t_0).
 \end{eqnarray}
 Of these, only the transverse ($D_x$) part of the hyperfine interaction picks up a time dependence.    Before we time transform the remaining two interaction Hamiltonians, we first spatially transform the conduction electron operators in the presence of the magnetic field necessary for the NMR experiment.  Then we rewrite ${\cal H}_{\rm e, cond}$ in terms of the spatially-transformed conduction electron fields,
 \begin{eqnarray}
 {\cal H}_{\rm e, cond}\rightarrow\tilde{\cal H}_{\rm e, cond}&=&\sum_{i,q,j}g_{{\rm L},j}\sum_{n_j=0}^{\infty}\int\frac{dk_{j,||}}{2\pi}\Bigl(\nu_{i,q,j}\tilde{\psi}^{\dag}_{j,n_j,\sigma}(k_{j,||})b_{i,q,\sigma}+H.c.\Bigr),
 \end{eqnarray}
 which after time-transformation with respect to ${\cal H}_{\rm e,0}$ and $\tilde{\cal H}_{\rm cond,0}$ becomes
 \begin{eqnarray}
 \tilde{\cal H}_{{\rm e,cond},{\cal H}_{\rm e,0}+\tilde{\cal H}_{\rm cond,0}}(t)&=&\sum_{i,q,j}g_{{\rm L}, j}\sum_{n_j=0}^{\infty}\int\frac{dk_{j,||}}{2\pi}\biggl(\nu_{i,q,j}e^{{\rm i}[\varepsilon_j(n_j,k_{j,||})-\varepsilon_q+\sigma(\omega_{\rm e}-\tilde{\omega}'_{j,\rm e})/2](t-t_0)}\nonumber\\& &\hskip50pt\tilde{\psi}^{\dag}_{j,n_j,\sigma}(k_{j,||},t_0)b_{i,q,\sigma}(t_0)
 +H.c.\biggr).
 \end{eqnarray}
 The most important Hamiltonian for the Knight shift in a superconductor is the pairing interaction ${\cal H}_{sc}$, which is position space was written in Equation 10.  Since in a Knight shift measurement, the experimenter first measures the Knight shift in the applied field ${\bm H}(t)$ and hence the induction ${\bm B}(t)=\mu_0{\bm H}(t)$ while the superconductor is in its normal (metallic) state, and then cools the material through its superconducting transition at $T_c(H)$, it is clear that the correct formulation for the superconducting pairing interaction must be in momentum space, and more precisely, to account for the pairing of the electrons (or holes) while they are in Landau orbits in the normal state. We therefore first rewrite ${\cal H}_{\rm sc}$ in a  fully spatially-transformed magnetic-induction-quantized form that allows for different pairing interactions, such as those giving rise to various types of spin-singlet and spin-triplet superconductors arising from a multiple-band metal.  As a start to understand the orbital motion of the paired superconducting electrons (or holes), we first assume the standard approximation for the evaluation of the upper critical field $H_{c2}$ that the  paired particles of combined charge $2{\rm e}$ move together in  Landau levels \cite{Scharnberg1980,Scharnberg1985,Werthamer1966,Klemm1975}.  For a BCS superconductor for which $V_{j,j';\sigma,\sigma'}({\bm k}_j-{\bm k}'_{j'})=-V_0\delta_{j,j'}\delta_{\sigma,-\sigma'}$, there is no need to transform the wave vector dependence of the pairing interaction due to the Landau orbits formed by the strong applied field \cite{Scharnberg1980}.  Such pairing interactions will be considered elsewhere.  Thus, we begin by considering only the simplest case  of isotropic intraband pairing of equivalent strength in all of the bands, which after spatial transformation due to the magnetic induction  may be written as
 \begin{eqnarray}
 \tilde{H}_{\rm sc,0}&=&-V_0\sum_{\sigma}\sum_jg^2_{{\rm L},j}\sum_{n_j,n'_j=0}^{\infty}\int\frac{dk_{j,||}}{2\pi}\int\frac{dk'_{j,||}}{2\pi}\tilde{\psi}^{\dag}_{j,n_j,\sigma}(k_{j,||})\tilde{\psi}^{\dag}_{j,n_j,-\sigma}(-k_{j,||})\tilde{\psi}^{}_{j,n'_j,-\sigma}(-k'_{j,||})\tilde{\psi}^{}_{j,n'_j,\sigma}(k'_{j,||}),\nonumber\\
 \end{eqnarray}
 and transforming this in time using ${\cal H}_{\rm cond,0}$, we have
 \begin{eqnarray}
 \tilde{H}_{\rm sc,0,\tilde{\cal H}_{\rm cond,0}}(t)&=&-V_0\sum_{\sigma}\sum_jg^2_{{\rm L},j}\sum_{n_j,n'_j=0}^{\infty}\int\frac{dk_{j,||}}{2\pi}\int\frac{dk'_{j,||}}{2\pi}\tilde{\psi}^{\dag}_{j,n_j,\sigma}(k_{j,||},t_0)\tilde{\psi}^{\dag}_{j,n_j,-\sigma}(-k_{j,||},t_0)\nonumber\\
 & &\hskip80pt\times\tilde{\psi}^{}_{j,n'_j,-\sigma}(-k'_{j,||},t_0)\tilde{\psi}^{}_{j,n'_j,\sigma}(k'_{j,||},t_0),
 \end{eqnarray}
 which is independent of $t$.
 \section{Dyson's Equations for the Green functions}
 For a system with continuous position variables ${\bm r}$, the contour $C$ Dyson equation for an adiabatic time-dependent interaction for which $C_{\rm int}\rightarrow C$ can be written as \cite{Haug2008}
 \begin{eqnarray}
 G(1,1')&=&G_0(1,1')+\int d^3{\bm r}_2\int_Cd\tau_2G_0(1,2)U(2)G(2,1')\nonumber\\
 & &+\int d^3{\bm r}_2\int d^3{\bm r}_3\int_Cd\tau_2\int_Cd\tau_3G_0(1,2)\Sigma(2,3)G(3,1'),
 \end{eqnarray}
 where $\Sigma$ is the self-energy and $U$ is an external field.
 By carefully keeping the order of the times in going about the contour $C$, one can analytically continue the integrals off the real axis to the real axis. We first need to define the retarded and advanced Green functions, which are
 \begin{eqnarray}
 G^{\rm r}(1,1')&=&\Theta(t_1-t_{1'})[G^{>}(1,1')-G^{<}(1,1')],\\
 G^{\rm a}(1,1')&=&\Theta(t_{1'}-t_1)[G^{<}(1,1')-G^{>}(1,1')].
 \end{eqnarray}
  Then, letting $\int_Cd\tau_2G_0(1,2)G(2,1')$ be represented by $C=AB$, one can analytically continue the appropriate contour-ordered Green function components on the real axis, so that
  \begin{eqnarray}
  C^{<}&=&A^{\rm r}B^{<}+A^{<}B^{\rm a},\\
  C^{>}&=&A^{\rm r}B^{>}+A^{>}B^{\rm a},
  \end{eqnarray}
  where the integration $\int_Cd\tau_2\rightarrow\int_{-\infty}^{\infty}dt_2$.
  Similarly, by representing the double contour integral $\int_Cd\tau_2\int_Cd\tau_3G_0(1,2)\Sigma(2,3)G(3,1')$ by $D=ABC$, one can analytically continue these contour integration paths to the real axis, obtaining\cite{Haug2008}
  \begin{eqnarray}
  D^{<}&=&A^{\rm r}B^{\rm r}C^{<}+A^{\rm r}B^{<}C^{\rm a}+A^{<}B^{\rm a}C^{\rm a},\\
  D^{>}&=&A^{\rm r}B^{\rm r}C^{>}+A^{\rm r}B^{>}C^{\rm a}+A^{>}B^{\rm a}C^{\rm a}.
  \end{eqnarray}
  We then implement the three Dyson equations for the nuclear, local orbital electron, and conduction electron Green functions.  We first consider the Dyson equation for the nuclear Green function.  In this case, there are two terms to consider:  the external field $U_{\rm n}$ given by Equation 56, and the hyperfine interaction given by Equation 59.  However, the hyperfine interaction does not involve two different times, as in the self-energy, which is analogous to the exchange interaction in the electron gas with electron-electron Coulomb interactions.  It is instead analogous to the direct interaction with a fermion loop, but in this case, the fermion loop is for the local orbital electrons.  As first shown by Hall and Klemm, the leading self-energy diagrams for the Knight shift and the linewidth changes in a metal at $T=0$ are shown in Figure 3.

 \begin{figure}
  \center{\includegraphics[width=0.3\textwidth]{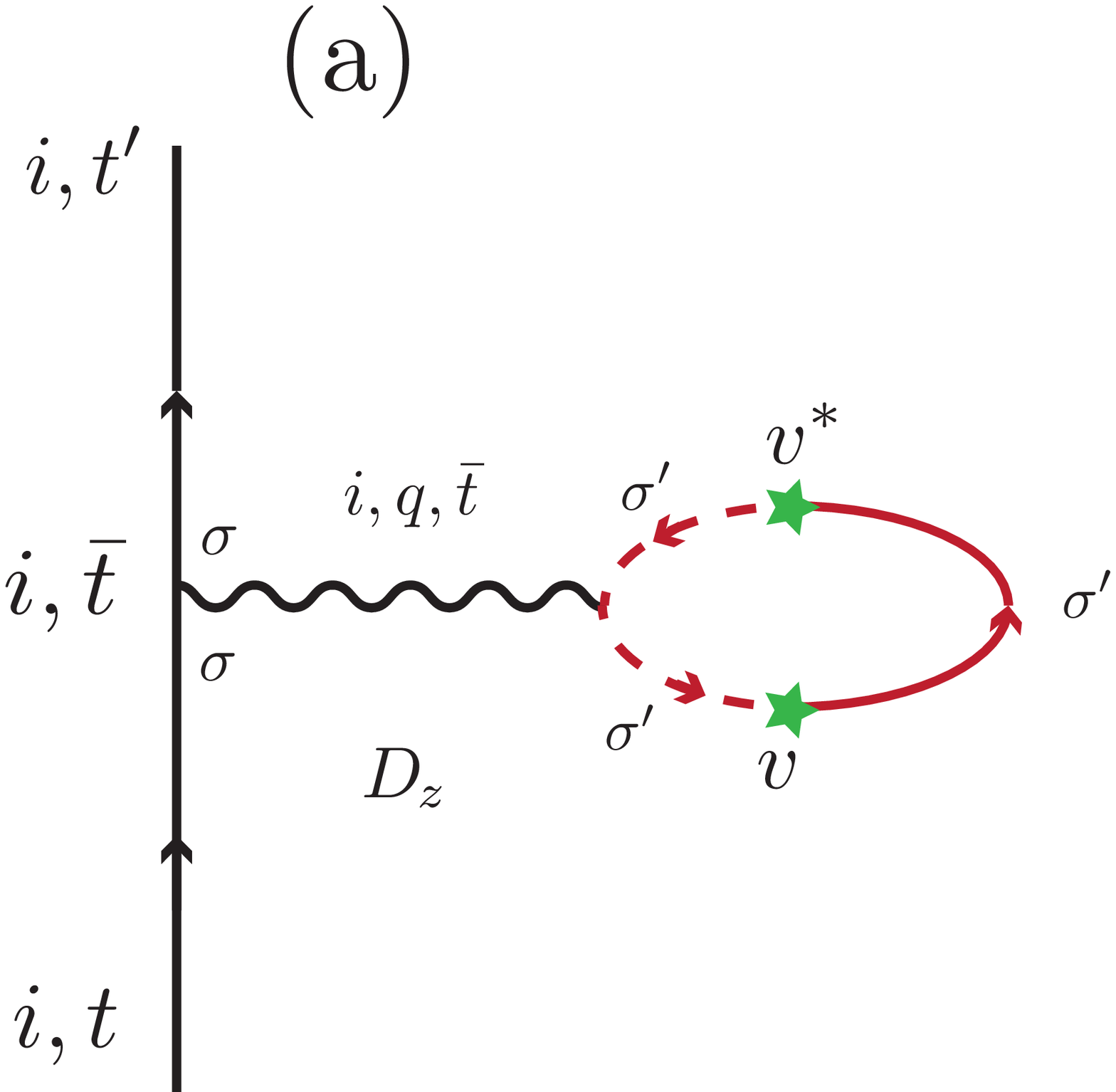}\hskip5pt\includegraphics[width=0.3\textwidth]{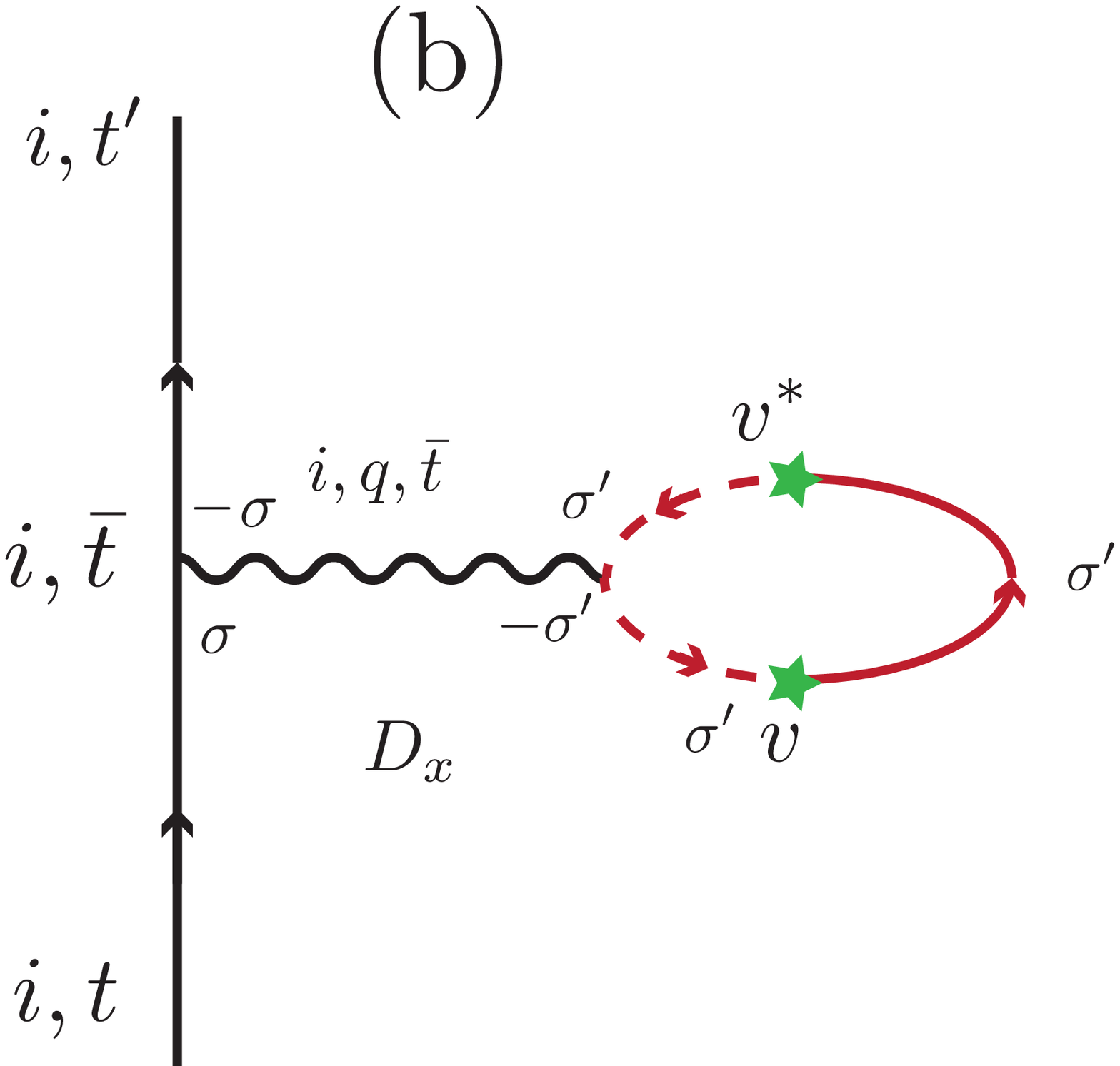}\hskip5pt
 \includegraphics[width=0.3\textwidth]{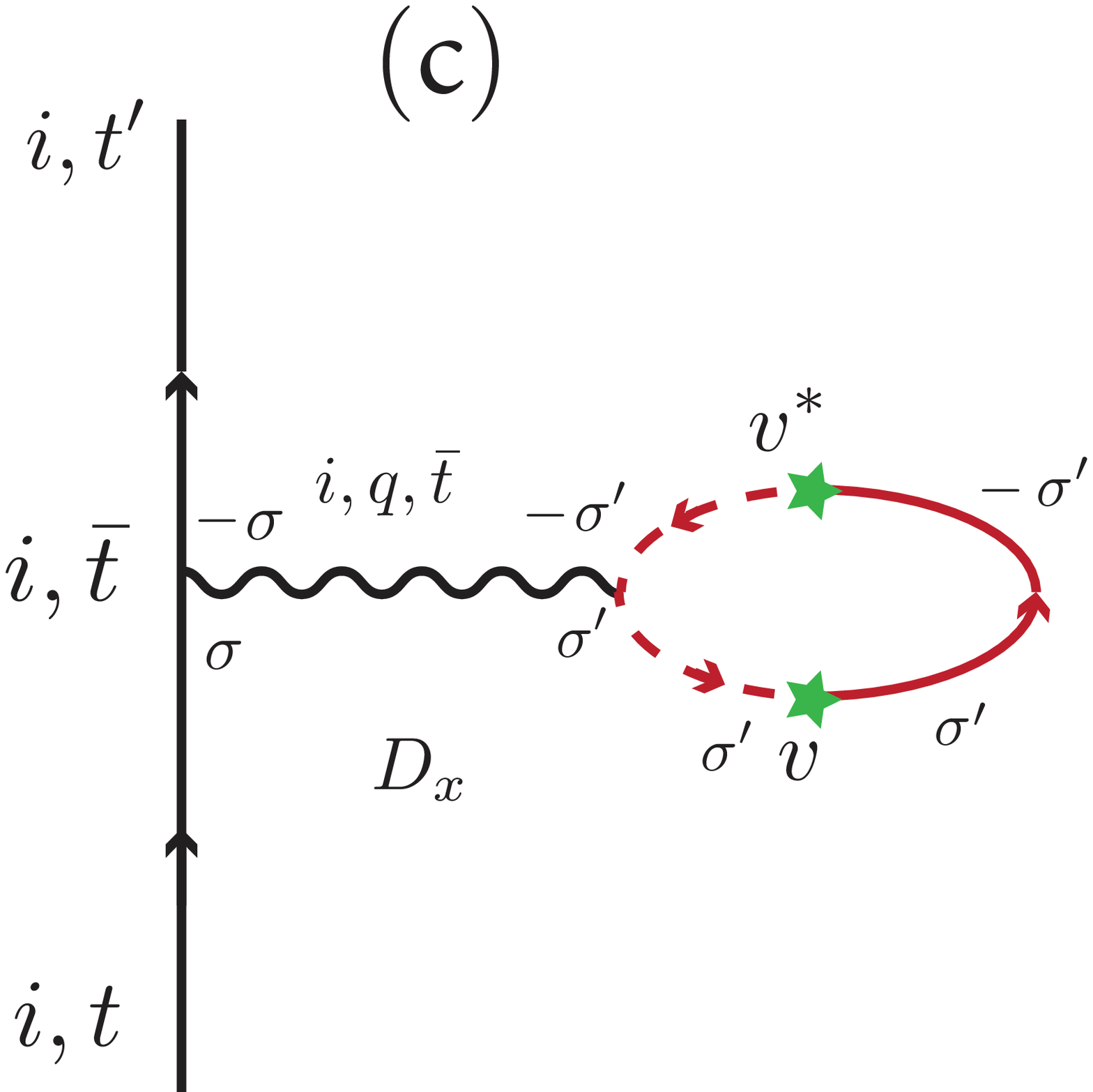}
 \caption{Hall-Klemm diagrams for the Knight shift linewidth changes at $T=0$.  The vertical solid lines on the left are the nuclear $G_{\rm n}(1,1')$, the wiggly horizontal line represents the hyperfine interaction, the dashed curves represent $G_{\rm e}$, the stars represent the excitation from the local orbitals to the conduction band, and the solid counterclockwise arrowed curves represent the conduction electron $G_{\rm cond}$.  (a) represents the leading Knight shift contribution arising from $D_z$.  (b) and (c) represent the two leading contributions to the linewidth changes \cite{Hall2016}.}}
 \end{figure}
 \section{Proposed Calculation of  $K(T)$ in the Normal and Superconducting States}
 \subsection{Gor'kov's Derivation of the Ginzburg-Landau Equations}
 Since the upper critical field has been obtained for anisotropic superconductors with a variety of pairing interactions \cite{Scharnberg1980,Scharnberg1985,Werthamer1966,Klemm1975} and also that the most rapid temperature variation,  a discontinuity in slope, of the conventional Knight shift in superconductors, occurs just at  the superconducting transition, it is evident that an extension of those upper critical field calculations to the Ginzburg-Landau regime just below $T_{c2}({\bm B}_0)$ can provide the crucial information for $K_S(T)$ in the superconducting state. We propose to extend the Hall-Klemm $T=0$ Knight shift calculation in the presence of a strong magnetic induction $B(t)$ into the superconducting state using an extension of the  microscopic derivation of the Ginzburg-Landau expression for the gap function as pioneered by Gor'kov\cite{Abrikosov1963,Gorkov1959} to include the time-dependent applied field.  That work was generalized for a general $V({\bm r}-{\bm r}')$ single-band pairing by Scharnberg and Klemm \cite{Scharnberg1980}.   In the superconducting state, we require the regular (conduction) and anomalous  Green functions
 \begin{eqnarray}
 G_{j,j';\sigma,\sigma'}(1,1')&=&-{\rm i}\langle T_{C}[\psi^{}_{j,\sigma,{\cal H}}(1)\psi^{\dag}_{j',\sigma',{\cal H}}(1')]\rangle,\\
 F_{j,j';\sigma,\sigma'}(1,1')&=&\langle T_C[\psi^{}_{j,\sigma,{\cal H}}(1)\psi^{}_{j',\sigma',{\cal H}}(1')]\rangle,\\
 F^{\dag}_{j,j'\sigma,\sigma'}(1,1')&=&\langle T_C[\psi^{\dag}_{j,\sigma,{\cal H}}(1)\psi^{\dag}_{j',\sigma',{\cal H}}(1')]\rangle,
 \end{eqnarray} and the gap function, which in real space for intraband pairing only is
 \begin{eqnarray}
 \Delta_{j;\sigma,\sigma'}({\bm r}_j,{\bm r}'_j)&=&V_{j}({\bm r}_j-{\bm r}'_j)\delta_{j,j'}F_{j,j';\sigma,\sigma'}(1,1')|_{t_{1}-t_{1'}=0+}.
 \end{eqnarray}
 As discussed in the next subsection, in order to include the temperature dependence of the normal state of the nuclei, the local orbital electrons, and the conduction electrons, we need to quantize the conduction electrons in momentum space and Landau orbits, as was done for the bare conduction electron Green functions in Equations 42 and 43.  Although this was never done in upper critical field calculations \cite{Klemm2012,Scharnberg1980,Zhang2014a,Scharnberg1985,Loerscher2013,Zhang2014b,Werthamer1966,Klemm1975}, the reasons given for not doing it were that impurities would broaden the levels, smearing out the Landau level spacings \cite{Scharnberg1980,Werthamer1966}.  However, with the present quality of some materials, that argument should be reexamined.  More important, in order to calculate the upper critical induction ${\bm B}_{c2}$, one requires the paired electrons (or holes) to be in Landau levels \cite{Klemm2012,Scharnberg1980,Zhang2014a,Scharnberg1985,Loerscher2013,Zhang2014b,Werthamer1966,Klemm1975}.  However, as discussed in the next section, it is not clear that electrons (or holes) will only pair with other electrons (or holes) in the same Landau orbit.  With multiple bands, an electron in one single-particle Landau level corresponding to one conduction band could in principle pair with another electron in a different Landau level corresponding to another band.  So one will have to make some assumptions about the pairing processes to simplify the calculations.  But to get a preliminary microscopic idea of how $K_S(T)$ picks up its $T$ dependence below $T_c$, we will first revisit the Gor'kov procedure for deriving the Ginzburg-Landau equations in real space.

 In the standard real-space finite temperature formalism, the  Gor'kov equations of motion generalized to include multiple ellipsoidally anisotropic bands and their Zeeman energies without interband pairing are\cite{Scharnberg1980}
 \begin{eqnarray}
 \Bigl({\rm i}\omega_n-\sum_{\nu=1}^3\frac{1}{2m_{j,\nu}}[{\nabla}_{j,\nu}/{\rm i}-{\rm e}{\bm A}_{j,\nu}({\bm r}_j)]^2+\mu-\sigma\omega'_{j,{\rm e}}/2\Bigr)G_{j,j';\sigma,\sigma'}({\bm r}_j,{\bm r}'_{j'},\omega_n)\nonumber\\
 +\sum_{\rho}\int d^3{\rm r}_j''\Delta^{}_{j;\sigma,\rho}({\bm r}_j,{\bm r}''_j)F^{\dag}_{j,j';\rho,\sigma'}({\bm r}_j'',{\bm r}'_{j'},\omega_n)&=&\delta_{\sigma,\sigma'}\delta_{j,j'}\delta^{(3)}({\bm r}_j-{\bm r}'_j),\nonumber\\
 & &\\
 \Bigl(-{\rm i}\omega_n-\sum_{\nu=1}^3\frac{1}{2m_{j,\nu}}[{\nabla}_{j,\nu}/{\rm i}+{\rm e}{\bm A}_{j,\nu}({\bm r}_j)]^2+\mu-\sigma\omega'_{j,{\rm e}}/2\Bigr)F^{\dag}_{j,j';\sigma,\sigma'}({\bm r}_j,{\bm r}'_j,\omega_n)\nonumber\\ +\sum_{\rho}\int d^3{\bm r}''_j\Delta^{*}_{j;\sigma,\rho}({\bm r}_j,{\bm r}''_j)G_{j,j';\rho,\sigma'}({\bm r}_j'',{\bm r}'_j,\omega_n)&=&0,
\end{eqnarray}
 where $\omega_n=(2n+1)\pi/\beta$ is the fermion Matsubara frequency.
 Letting $G^{(0)}_{j,j';\sigma,\sigma'}({\bm r}_j,{\bm r}_{j'}',\omega_n)$ be the solution for the $G$ function in the normal state with $\Delta=0$,  one can rewrite the above equations for $G$ and $F^{\dag}$ in the finite temperature formalism as
 \begin{eqnarray}
 G_{j,j';\sigma,\sigma'}({\bm r}_j,{\bm r}'_{j'},\omega_n)&=&G_{j,j;\sigma,\sigma'}^{(0)}({\bm r}_j,{\bm r}'_{j'},\omega_n)\delta_{j,j'}\delta_{\sigma,\sigma'}\nonumber\\
 & &-\sum_{\rho}\int d^3{\bm r}_j''\int d^3{\bm r}_j'''G^{(0)}_{j,j;\sigma,\sigma}({\bm r}_j,{\bm r}_j''',\omega_n)\Delta_{j,\rho,\sigma}({\bm r}'''_j,{\bm r}''_j)\nonumber\\
 & &\times\int d^3{\bm \xi}_j\int d^3{\bm \xi}'_j\sum_{\rho'}G^{(0)}_{j,j;\sigma',\sigma'}({\bm r}'_j,{\bm \xi}'_j,-\omega_n)\Delta^{*}_{j,\rho',\sigma'}({\bm \xi}_j,{\bm\xi}'_j)G^{(0)}_{j,\rho,\rho'}({\bm r}''_j,{\bm \xi}_j,\omega_n),\nonumber\\
 \end{eqnarray} to order $\Delta^2$. One can then substitute this in the equation for $F^{\dag}$, multiply by the pairing interaction, and obtain a self-consistent equation for $\Delta$ to order $\Delta^3$ \cite{Abrikosov1963,Gorkov1959}.
   The coefficients of the  two terms proportional to $\Delta$  define the upper critical induction ${\bm B}_{c2}$ \cite{Scharnberg1980,Zhang2014a,Scharnberg1985,Loerscher2013,Zhang2014b,Werthamer1966,Klemm1975}.  Functionally integrating the cubic equation  with respect to $\Delta_{j,\sigma,\sigma'}({\bm r}_j,{\bm r}'_j)$ and by neglecting the field dependence of the resulting term proportional to $|\Delta|^4$, one can obtain the generalized Ginzburg-Landau free energy.

We note that even if one completely neglects the field dependence of the term of order $\Delta^3$ in the Gor'kov expansion of $F^{\dag}$ for $\Delta({\bm r})$, it is easy to see that this procedure will lead to the following phenomenological general result:
\begin{eqnarray}
K_S(T)&=&
a({\bm B}_0)-b({\bm B}_0)|\Delta({\bm B}_0,T)|^2,
\end{eqnarray}
where  $a$ and $b$ strongly depend upon the magnitude and direction of ${\bm B}_0$, but not much upon $T$, and $2|\Delta({\bm B}_0,T)|$ is the effective superconducting gap in the Ginzburg-Landau regime.  This simple result includes the pairing in all of the bands, which couple together to give one effective $T_{c2}({\bm B}_0)$, below  which $|\Delta({\bm B}_0,T)|^2\propto [T_{c2}({\bm B}_0)-T]$. It remains to be seen if this form could be generalized to the full BCS superconducting gap $|\Delta({\bm B}_0,T)|$ temperature dependence, which saturates at low $T$ values. If so, it could lead to a quantitative theory of the Knight shift that would be valid for essentially any type of superconductor involving Cooper pairing. Hence, a proper calculation of $a({\bm B}_0)$ and $b({\bm B}_0)$  can provide a microscopic understanding of the behavior for the $^{63}$CuO$_2$  $K_S(T)$ for ${\bm B}_0$ parallel and normal to the layers of YBa$_2$Cu$_3$O$_{7-\delta}$, which was described by Slichter as ``fortuitous''\cite{Slichter1999}.  It could in principle explain the small or vanishing $b$ term in Sr$_2$RuO$_4$, at least for the field normal to the layers, for which Landau level formation would be highly restricted on two of the Fermi surfaces.
\subsection{High-Field Solution for an Anisotropic, Multiband Type-II BCS Superconductor}
 More important, we note that a major simplification of the Keldysh contour procedure can be made by first taking the mean-field approximation of the BCS pairing interaction represented in momentum space  by Equation 63.  We write the mean-field gap (or isotropic order parameter) for singlet pairing in band $j$ as
\begin{eqnarray}
\Delta_{j,-\sigma,\sigma}&=&V_0g_{{\rm L},j}\sum_{n_j=0}^{\infty}\int\frac{dk_{j_{||}}}{2\pi}\langle\tilde{\psi}_{j,n_j,-\sigma}(-k_{j,||})\tilde{\psi}_{j,n_j,\sigma}(k_{j,||})\rangle,
\end{eqnarray}
where the expectation value is in the grand canonical ensemble,
so that the mean-field effective Hamiltonian for the conduction electrons in the superconducting state becomes
\begin{eqnarray}
\tilde{\cal H}_{\rm sc,cond}&=&\sum_{j,\sigma}g_{{\rm L},j}\sum_{n_j=0}^{\infty}\int\frac{dk_{j,||}}{2\pi}\biggl(\tilde{\psi}^{\dag}_{j,n_j,\sigma}(k_{j,||})[\varepsilon_j(n_j,k_{j,||})-\mu_{\rm cond,cp}-\sigma\tilde{\omega}'_{j,{\rm e}}/2]\tilde{\psi}^{}_{j,n_j,\sigma}(k_{j,||})\nonumber\\
& &+\Bigl[\tilde{\psi}^{\dag}_{j,n_j,\sigma}(k_{j,||})\tilde{\psi}^{\dag}_{j,n_j,-\sigma}(-k_{j,||})\Delta_{j,-\sigma,\sigma}+H.c.\Bigr]\biggr).
\end{eqnarray}
where we have included the chemical potential of the conduction electrons.  Note that we assume the total momentum of the paired electrons (or holes) is zero, as both are assumed to be on opposite sides of the same Landau orbit, and have opposite momenta in the direction normal to the plane of the Landau orbits.  This effective quadratic Hamiltonian can then be diagonalized by a standard Bogoliubov-Valatin transformation \cite{Rickayzen1969}, letting
\begin{eqnarray}
\tilde{\psi}^{}_{j,n_j,\uparrow}(k_{j,||})&=&u^{}_{j,n_j,k_{j,||}}\gamma^{}_{j,n_j,\uparrow}(k_{j,||})+v^{}_{j,n_j,k_{j,||}}\gamma^{\dag}_{j,n_j,\downarrow}(k_{j,||}),\\
\tilde{\psi}^{\dag}_{j,n_j,\downarrow}(-k_{j,||})&=&-v^{*}_{j,n_j,k_{j,||}}\gamma^{}_{j,n_j,\uparrow}(k_{j,||})+u^{*}_{j,n_j,k_{j,||}}\gamma^{\dag}_{j,n_j,\downarrow}(k_{j,||}),
\end{eqnarray}
where we require the $\gamma$ operators to obey independent fermion statistics.  Using the standard transformation procedure to eliminate the off-diagonal terms, we then obtain
\begin{eqnarray}
|u^{}_{j,n_j,k_{j,||}}|^2&=&\frac{1}{2}\Bigl[1+\Bigl(\frac{\varepsilon_j(n_j,k_{j,||})-\mu_{\rm cond,cp}}{E_j(n_j,k_{j,||})}\Bigr)\Bigr],\\
|v^{}_{j,n_j,k_{j,||}}|^2&=&\frac{1}{2}\Bigl[1-\Bigr(\frac{\varepsilon_j(n_j,k_{j,||})-\mu_{\rm cond,cp}}{E_j(n_j,k_{j,||})}\Bigr)\Bigr],
\end{eqnarray}
and the diagonalized superconducting Hamiltonian becomes
\begin{eqnarray}
\tilde{\cal H}_{\rm sc,cond}&\rightarrow&\sum_{j,\sigma}g_{{\rm L},j}\sum_{n_j=0}^{\infty}\int\frac{dk_{j,||}}{2\pi}\gamma^{\dag}_{j,n_j,\sigma}(k_{j,||})\gamma^{}_{j,n_j,\sigma}(k_{j,||})\Bigl[E_j(n_j,k_{j,||})+\sigma\tilde{\omega}'_{j,{\rm e}}/2\Bigr],\\
E_j(n_j,k_{j,||})&=&\sqrt{[\varepsilon_j(n_j,k_{j,||})-\mu_{\rm cond,cp}]^2+|\Delta_{j}|^2},
\end{eqnarray}
where $|\Delta_j|^2=\Delta_{j,-\sigma,\sigma}\Delta^{\dag}_{j,-\sigma,\sigma}$ is positive definite for each $j$ value.
We note that the quasiparticle dispersions in $\tilde{\cal H}_{\rm sc,cond}$ are nearly identical to the BCS quasiparticle dispersions, as they do indeed have a real energy gap $2|\Delta|$, but there is in addition an effective Zeeman term arising from the difference in the spin up and spin down quasiparticle energies, leading to a magnetic gap function.  Thus, the self-consistent expression from Equation 80 for $\Delta_{j,-\sigma,\sigma}$ becomes
\begin{eqnarray}
\Delta_{j,\downarrow,\uparrow}&=&-V_0g_{{\rm L},j}\sum_{\sigma=\pm}\sum_{n_j=0}^{\infty}\int\frac{dk_{j,||}}{2\pi}\frac{\Delta_{j,\downarrow,\uparrow}}{E_j(n_j,k_{j,||})}\biggl(\frac{1}{e^{\beta[E_j(n_j,k_{j,||})+\sigma\tilde{\omega}'_{j,{\rm e}}/2]}+1}-\frac{1}{2}\biggr),
\end{eqnarray}
which, combined with Equation 86, explicitly demonstrates the presence of the superconducting gap $\Delta_j$ in each band that is involved in $K_S(T)$ in the superconducting state.  Thus, it is clear that the effective or phenomenological Equation 79 mentioned in the abstract for $K_S(T)$ applies in the mixed state of a type-II superconductor, not just in the Ginzburg-Landau region.  However, by quantizing the superconducting order parameter at a finite induction strength ${\bm B}_0$, both the Landau orbits and the Zeeman interaction can greatly affect its ${\bm B}_0$ dependence, and the Landau orbits in particular can be distinctly different for layered compounds with ${\bm B}_0$ parallel or perpendicular to the layers, especially at large induction strengths, as first noted in experiments on  YBa$_2$Cu$_3$O$_{7-\delta}$ \cite{Barrett1990,Slichter1999}.

 The road ahead to construct the first microscopic theory of the Knight shift in a superconductor of any type is now clear.  The conduction electrons must be quantized in Landau orbits, and this can be done for any number of ellipsoidally anisotropic electron or hole bands, as outlined above.  The procedure will  be extended for our model of multiple ellipsoidal bands with the Zeeman couplings and time-dependent Zeeman couplings in each band to construct the Bogoliubov-Valatin transformed contour $G$  functions. To do this properly, one needs to apply those transformations presented in Equations 82 and 83 to the time-dependent Zeeman interaction on the conduction electrons in Equations 44 and also to the Anderson-like interaction in Equation 61 that removes a local orbital electron and places it in the superconducting state and {\em vice versa}. This will cause Equations 44, 55, 61, and 62 to be rewritten in terms of the quasiparticle operators $\gamma^{}_{j,n_j,\sigma}(k_{j,||})$ and $\gamma^{\dag}_{j,n_j,\sigma}(k_{j,||})$, and will modify Equation 58. Then, the Keldysh contour method can be used to perform a microscopic theory of $K_S(T)$ in the superconducting state of an anisotropic, multiband BCS superconductor.  Since the conduction electrons are transformed into non-interacting quasiparticles in the superconducting state, the self-energy $\Sigma(2,3)$ in Dyson's equation will only apply to the orbital electrons via the Hubbard interaction $U_q$.  All other interactions reduce to effective external fields. After a detailed microscopic evaluation of $K_S(T)$ using the contour-extended version of the diagram pictured in Figure 3(a), special attention will be directed at the conditions for a near vanishing of $b({\bm B}_0)$, which could lead to a $T$-independent $K_S(T)$, even for a ``conventional'' superconductor.  The linewidth changes can be evaluated in the superconducting state from the contour-extended versions of the diagrams pictured in Figures 3(b,c).  Eventually, other superconducting pairing symmetries could also be studied with this technique, although the pairing interaction would have to be transformed as above, including the Landau orbits.  Eventually, this could be done for charge-density and spin-density wave systems, for which no theory of the Knight shift is presently available.  We note that 2$H$-TaS$_2$ has a nodal charge-density wave below 75 K, with a presumably $s$-wave superconducting state entering below 0.6 K \cite{Klemm2012,Klemm2015}, which is very similar to the complex situation in the high-temperature superconductor Bi$_2$Sr$_2$CaCu$_2$O$_{8+\delta}$, in which the nodal pseudogap (probably charge-density wave) regions and isotropic $s$-wave superconducting regions break up into spatial domains \cite{Zhong2016}.  These results are consistent with previous $c$-axis twist Josephson junction experiments on that material \cite{Li1999}.  Although the NMR linewidths in that material are too broad to perform Knight shift measurements, they could be done on other materials, such as the dichalcogenides, and also in improved YBa$_2$Cu$_3$O$_{7-\delta}$ samples.
 \section{Discussion and Conclusions}
 We have outlined a procedure to obtain a microscopic theory of the Knight shift in an anisotropic Type-II superconductor.  This was based upon the Hall-Klemm microscopic model of the effect at $T=0$ \cite{Hall2016}, for which multiple anisotropic conduction bands of ellipsoidal shapes were included.  We considered the simplest magnetic resonance case of ${\bm B}(t)={\bm B}_0+{\bm B}_1(t)$ with $|{\bm B}_1|\ll|{\bm B}_0|$ and ${\bm B}_1\cdot{\bm B}_0=0$ with ${\bm B}_1(t)$ oscillating at a single frequency $\omega_0$.  For this simple case, the time changes to the system are adiabatic, so that the interaction Keldysh contour  $C_{\rm int}$ effectively coincides with contour $C$ depicted in Figure 1, and the integrations can be analytically continued onto the real axis.  The procedure can effectively treat any nuclear spin value $I$.  The conduction electrons (or holes) were quantized in Landau orbits in the applied field in the normal state, and the Hamiltonian for a generalized anisotropic, multiband BCS type-II superconductor was  diagonalized, allowing  for a full treatment of the superconducting state.

 We emphasize that by quantizing the superconducting order parameter in the presence of a strong time-independent magnetic induction ${\bm B}_0$, the energy spacings of the Landau orbits can depend strongly upon the direction of ${\bm B}_0$.  At very weak ${\bm B}_0$ values, the Landau levels primarily give rise to  overall anisotropic constant backgrounds of $K(T)$ and $K_S(0)$, with $K_S(T)$ being predominantly governed by the anisotropic Zeeman interactions and $D_{z}$.  But for sufficiently strong ${\bm B}_0$ values in anisotropic materials with layered or quasi-two-dimensional anisotropy, the spacings between the Landau energy levels depends strongly upon the direction of ${\bm B}_0$, so that $K_S(T)$ could become independent of $T$ for $T\le T_c$, as first observed for ${\bm B}_0||\hat{\bm c}$ in YBa$_2$Cu$_3$O$_{7-\delta}$ \cite{Barrett1990,Slichter1999}.  Such behavior could also arise for quasi-one-dimensional materials in all ${\bm B}_0$ directions, although to different degrees for ${\bm B}_0$ parallel and perpendicular to the most conducting crystal direction.

 Since the crucial interaction for the Knight shift is the hyperfine interaction between the probed nuclei and their surrounding orbital electrons, the symmetry of this interaction can be very important. Generally, the hyperfine interaction can arise from the electrons in any of the orbital levels.  For $s$-orbitals, the Fermi contact term is important,  but the induced-dipole induced-dipole interactions can arise from the nucleus of any spin for any spin $I\ge 1/2$ and its surrounding electrons in any orbital, and induced-quadrupole induced-quadrupole and higher order interactions can also occur for certain orbitals and nuclear spin values.  In the Hall-Klemm model \cite{Hall2016}, the hyperfine interaction crucial for the Knight shift was taken to be diagonal in the spin representations of a lattice with  tetragonal symmetry $D_x=D_y\ne D_z$.  In that simple model, the $T=0$ results indicated that the Knight shift arose from $D_z$, and the line width was modified by $D_x=D_y$.  In more realistic examples of correlated and anisotropic materials, the hyperfine interaction  would be represented by a symmetric matrix unless time-reversal symmetry-breaking interactions were present.  Such matrices can be diagonalized by a set of rotations, but in complicated cases the quantization axes would not necessarily be the same as for the overall crystal structure.  Such complications would mix the Knight shift and its linewidth, depending upon the direction of ${\bm B}_0$.

 As noted previously, in first quantization, an isolated nuclear spin wave function in an NMR experiment was found to have the form
 \begin{eqnarray}
 |I,m_I\rangle(t)&=&e^{{\rm i}m_I\omega_0t}\sum_{m'_I=-I}^IC_{m'_I}^{m_I}e^{{\rm i}m_I'\Gamma_{\rm n}t},
 \end{eqnarray}
 where $\Gamma_{\rm n}=[(\omega_0-\omega_{\rm n})^2+\Omega_{\rm n}^2]^{1/2}$ is the nuclear resonance function and the constants $C_{m_I'}^{m_I}$ depend upon the initial conditions \cite{Hall2016}.  Those authors found this form to hold for $I=1/2,1, 3/2$, and in second quantization, up to $I=2$, so it is likely to hold for arbitrary $I$.  In the adiabatic regime, we have $\omega_0\ll\omega_{\rm n}$\cite{Berry1984,Griffiths2005}, so that there will be a manifold of geometrical phases that will arise with higher $I$ values.

 We remark that it is possible to generalize this treatment to more complicated ${\bm B}_1(t)$ functions, such as a periodic function of square-wave or triangle-wave shape.  This can be represented as a Fourier series, but if the primary angular frequency is $\omega_0$, terms of higher multiples $n$ of $\omega_0$ can be present, some of which would violate the adiabatic requirement that they be much smaller than the Zeeman energy spacings.  Hence, this experiment would make some amount of  non-adiabatic changes that could drive the system out of thermal equilibrium, and the two contours discussed above would not coincide, greatly complicating the analysis.



\begin{thebibliography}{999}
 \bibitem[Knight(1949)]{Knight1949}
  Knight, W. D. Nuclear Magnetic Resonance Shift in Metals. {\em Phys. Rev.} {\bf 1949}, {\em 76}, 1259-1260. DOI: 10.1103/PhysRev.76.1259.2
\bibitem[Knight(1956)]{Knight1956}
  Knight, W. D., Androes, G. M., and Hammond, R. H. Nuclear Magnetic Resonance in a Superconductor. {\em Phys. Rev.} {\bf 1956}, {\em 104}, 852-853. DOI: 10.1103/PhysRev.104.852
\bibitem[Reif(1956)]{Reif1956}
  Reif, F. Observation of Nuclear Magnetic Resonance in Superconducting Mercury. {\em Phys. Rev.} {\bf 1956}, {\em 102}, 1417-1418. DOI: 10.1103/PhysRev.102.1417
  \bibitem[Reif(1957)]{Reif1957}
  Reif, F. The Study of Superconducting Hg by Nuclear Magnetic Resonance Techniques. {\em Phys. Rev.} {\bf 1957}, {\em 106}, 208-221. DOI: 10.1103/PhysRev.106.208
  \bibitem[Yosida(1958)]{Yosida1958}
  Yosida, K. Paramagnetic Susceptibility in Superconductors. {\em Phys. Rev.} {\bf 1958}, {\em 110}, 769-770. DOI: 10.1103/PhysRev.110.769
  \bibitem[Gladstone(1969)]{Gladstone1969}
Gladstone, G., Jensen, M. A. and Schrieffer, J. R.  Superconductivity in the Transition Metals, in Parks, R. D. {\em Superconductivity}, Marcel Dekker, Inc., New York, 1969, pp. 801-803. ISBN: 978-0824715212.
  \bibitem[Abrikosov(1962)]{Abrikosov1962}
  Abrikosov, A. A. and Gor'kov, L. P. Spin-Orbit Interaction and the Knight Shift in Superconductors. {\em Sov. Phys.--JETP} {\bf 1962}, {\em 15}, 752-57.
  \bibitem[Baek(2013)]{Baek2013}
  Baek, S.-H., Harnegea, L., Wurmehl, S., B{\"u}chner, B. and Grafe, H.-J. Anomalous Superconducting State in LiFeAs Implied by the $^{75}$As Knight Shift Measurement. {\em J. Phys.: Condens. Matter}, {\bf 2013}, {\em 25}, 162204.  DOI: 10.1088/0953-8984/25/16/162204.
  \bibitem[Kohori(2001)]{Kohori2001}
  Kohori, Y., Yamato, Y., Iwamoto, Y., Kohara, T., Bauer, E. D., Maple, M. B. and Sarrao, J. L. NMR and NQR Studies of the Heavy Fermion Superconductors CeTIn$_5$ (T = Co and Ir). {\em Phys. Rev. B}, {\bf 2001}, {\em 64}, 134526. DOI: 10.1103/Phys RevB.64.134526.
  \bibitem[Lee(2002)]{Lee2002}
  Lee, I. J., Brown, S. E., Clark, W. G., Strouse, M. J., Naughton, M. J., Kang, W. and Chaikin, P. M. Triplet Superconductivity in an Organic Superconductor Probed by NMR Knight Shift. {\em Phys. Rev. Lett.}, {\bf 2002}, {\em 88}, 017004. DOI: 10.1103/PhysRevLett.88.017004
  \bibitem[Lee(2003)]{Lee2003}
  Lee, I. J., Brown, S. E., Clark, W. G., Strouse, M. J., Naughton, M. J.,  Chaikin, P. M. and Brown, S. E. Evidence from $^{77}$Se Knight Shifts for Triplet Superconductivity in (TMTSF)$_2$PF$_6$. {\em Phys. Rev. B}, {\bf 2003}, {\em 68}, 092510. DOI: 10.1103/PhysRevB.68.092510.
  \bibitem[Michioka(2006)]{Michioka2006}
  Michioka, C., Ohta, H., Itoh, Y., Yoshimura, K., Kato, M., Sakurai, H., Takayama-Muromachi, E., Takada, K. and Sasaki, T. Knight Shift of Triangular Lattice Superconductor Na$_{0.35}$CoO$_2\cdot$1.3H$_2$O. {\em Physica B}, {\bf 2006}, 628-629. DOI: 10.1016/j.physb.2006.01.181.
  \bibitem[Kato(2006)]{Kato2006}
  Kato, M., Michioka, C., Waki, T., Itoh, Y., Yoshimura, K., Ishida, K., Sakurai, H., Takayama-Muromachi, E., Takada, K. and Sasaki, T. Possible Spin Triplet Superconductivity in Na$_x$CoO$_2\cdot y$H$_2$O:  $^{59}$Co NMR Studies. {\em J. Phys.: Condens. Matter}, {\bf 2006}, {\em 18}, 669-682. DOI: 10.1088/0953-8984/18/2/022.
  \bibitem[Sakurai(2015)]{Sakurai2015}
  Sakurai, H., Ihara, Y. and Takada, K. Superconductivity of Cobalt Oxide Hydrate Na$_x$(H$_3$O)$_z$CoO$_2\cdot y$H$_2$O. {\em Physica C} {\bf 2015}, 378-387.  DOI: 10.1016/j.physc.2015.02.010.
  \bibitem[Klemm(2000)]{Klemm2000}
  Klemm, R. A. Striking Similarities Between the Pseudogap Phenomenon in the Cuprates and in Layered Organic and Dichalcogenide Superconductors. {\em Physica C} {\bf 2000}, {\em 341-348}, 839-842. DOI:
  \bibitem[Chou(2004)]{Chou2004}
  Chou, F. C., Cho, J. H., Lee, P. A., Abel, E. T., Matan, K., and Lee, Y. S. Thermodynamic and transport measurements of superconducting  Na$_{0.3}$CoO$_2\cdot1.3$H$_2$O single crystals prepared bu electrochemical deintercalation. {\em Phys. Rev. Lett.} {\bf 2004}, {\em 92}, 157004.  DOI: 10.1103/PhysRevLett.92.157004.
  \bibitem[Klemm(2012)]{Klemm2012}
   Klemm, R. A. {\em Layered Superconductors Volume 1}, Oxford University Press, Oxford, UK, 2012. ISBN 978-0-19-959331-6.
 \bibitem[Scharnberg(1980)]{Scharnberg1980}
  Scharnberg, K. and Klemm, R. A. $P$-Wave Superconductors in Magnetic Fields. {\em Phys. Rev. B} {\bf 1980}, {\em 22}, 5233-44. DOI: 10.1103/PhysRevB.22.5233.
   \bibitem[Barrett(1990)]{Barrett1990}
  Barrett, S. E., Durand, D. J., Pennington, C. H., Slichter, C. P., Friedmann, T. A., Rice, J. P., and Ginsberg, D. M. $^{63}$Cu Knight Shifts in the Superconducting State of YBa$_2$Cu$_3$O$_{7-\delta}$ ($T_c=90$ K). {\em Phys. Rev. B}, {\bf 1990}, {\em 41}, 6283-6296. DOI: 10.1103/PhysRevB.41.6283
 \bibitem[Slichter(1999)]{Slichter1999}
  Slichter, C. P., The Knight Shift -- A Powerful Probe of Condensed-Matter Systems. {\em Phil. Mag. B}, {\bf 1999}, {\em 79}, 1253-1261. DOI: 10.1080/13642819908216968
  \bibitem[Fujiwara(1991)]{Fujiwara1991}
  Fujiwara, K., Kitaoka, Y., Ishida, K., Asayama, K., Shimakawa, Y., Manako, T. and Kubo, Y. NMR and NQR Studies of Superconductivity in Heavily Doped Tl$_2$Ba$_2$CuO$_{6+y}$ with a Single CuO$_2$ Plane.  {\em Physica C} {\bf 1991}, {\em 184}, 207-219.  DOI: 10.1016/0921-4534(91)90385-C.
  \bibitem[Zheng(1996)]{Zheng1996}
  Zheng, G. Q., Kitaoka, Y., Asayama, K., Hamada, K., Yamauchi, H. and Tanaka, S. NMR Study of Local Hole Distribution, Spin Fluctuation, and Superconductivity in Tl$_2$Ba$_2$Ca$_2$Cu$_3$O$_{10}$. {\em Physica C} {\bf 1996}, {\em 260}, 197-210. DOI: 10.1016/0921-4534(96)000092-5.
 \bibitem[Zheng(2003)]{Zheng2003}
  Zheng, G. Q., Sato, T., Kitaoka, Y., Fujita, M. and Yamada, K. Fermi-Liquid Ground State in the $n$-Type Pr$_{0.91}$LaCe$_{0.09}$CuO$_{4-y}$ Copper-Oxide Superconductor. {\em Phys. Rev. Lett.} {\bf 2003}, {\em 90}, 197005. DOI: 10.1103/PhysRevLett.90.197005.
  \bibitem[Kotegawa(2008)]{Kotegawa2008}
  Kotegawa, H., Masaki, S., Awai, Y., Tou, H., Mizuguchi, Y. and Takano Y. Evidence for Unconventional Superconductivity in Arsenic-Free Iron-Based Superconductor FeSe:  A $^{77}$Se-NMR Study. {\em J. Phys. Soc. Jpn.} {\bf 2008}, {\em 77}, 113703. DOI: 10.1143/JPSJ.77.113703.
 \bibitem[Ishida(1998)]{Ishida1998}
  Ishida, K., Mukuda, H., Kitaoka, Y., Asayama, K., Mao, Z. Q.,  Mori, Y. and Maeno, Y.  Spin-Triplet Superconductivity in Sr$_2$RuO$_4$ Identified by $^{17}$O Knight Shift. {\em Nature}, {\bf 1998}, {\em 396}, 658-660. DOI: 10.1038/25315.
 \bibitem[Ishida(2001)]{Ishida2001}
   Ishida, K., Mukuda, H., Kitaoka, Y.,  Mao, Z. Q.,  Fukazawa, H. and Maeno, Y. Ru NMR Probe of the Spin Susectibility in the Superconducting State of Sr$_2$RuO$_4$.{\em Phys. Rev. B}, {\bf 2001}, {\em 63}, 060507. DOI: 10.1103/PhysRevB.63.060507.
  \bibitem[Mackenzie(2003)]{Mackenzie2003}
   Mackenzie, A. P. and Maeno, Y. The Superconductivity of Sr$_2$RuO$_4$ and the Physics of Spin-Triplet Pairing. {\em Rev. Mod. Phys.} {\bf 2003}, {\em 25}, 657-712. DOI: 10.1103/RevModPhys.75.657.
   \bibitem[Duffy(2000)]{Duffy2000}
   Duffy, J. A., Hayden, S. M., Maeno, Y., Mao, Z., Kulda, J. and McIntyre, G. J.  Polarized-Neutron Scattering Study of the Cooper-Pair Moment in Sr$_2$RuO$_4$. {\em Phys. Rev. Lett.}, {\bf 2000}, {\em 85}, 5412-5415.  DOI: 10.1103/PhyRevLett.85.5412.
   \bibitem[Deguchi(2002)]{Deguchi2002}
   Deguchi, K., Tanatar, M. A., Mao, Z. Q., Ishiguro, T. and Maeno, Y.  Superconducting Double Transiton and the Upper Critical Field Limit of Sr$_2$RuO$_4$ in Parallel Magnetic Fields. {\em J. Phys. Soc. Jpn.} {\bf 2002}, {\em 71}, 2839-2842. DOI: 10.1143/JPSJ.71.2839
   \bibitem[Kittaka(2009)]{Kittaka2009}
   Kittaka, S., Nakamura, T., Aono, Y. Yonezawa, S., Ishida, K. and Maeno, Y. Angular Dependence of the Upper Critical Field of Sr$_2$RuO$_4$. {\em Phys. Rev. B}, {\bf 2009}, {\em 80}, 174514. DOI: 10.1103/PhysRevB.80.174514.
   \bibitem[Machida(2008)]{Machida2008}
   Machida, K. and Ichioka, M. Magnetic Field Dependence of Low-Temperature Specific Heat in Sr$_2$RuO$_4$. {\em Phys. Rev. B}, {\bf 2008}, {\em 77}, 184515. DOI: 10.1103/PhysRevB.77.184515.
  \bibitem[Zhang(2014a)]{Zhang2014a}
   Zhang, J., L{\"o}rscher, C., Gu, Q. and Klemm, R. A. Is the Anisotropy of the Upper Critical Field of Sr$_2$RuO$_4$ Consistent with a Helical $p$-Wave State? {\em J. Phys.: Condens, Matter} {\bf 2014}, {\em 26}, 252201 DOI: 10.1088/0953-8984/26/25/252201.
   \bibitem[Annett(2007)]{Annett2007}
   Annett, J. F., Gy{\"o}rffy, B. L.,  Litak, G. and Wysoki{\'n}ski, K. I. Magnetic Field Inducted Rotation of the ${\bm d}$-Vector in Sr$_2$RuO$_4$. {\em Physica C} {\bf 2007}, {\em 460-2}, 995-996. DOI: 10.1016/j.physc.2007.03.377.
   \bibitem[Leggett(1975)]{Leggett1975}
   Leggett, A. J. A Theoretical Description of the New Phases of Liquid $^3$He. {\em Rev. Mod. Phys.} {\bf 1975}, {\em 47}, 331-414. DOI: 10.1103/RevModPhys.47.331.
   \bibitem[Rozbicki(2011)]{Rozbicki2011}
   Rozbicki, E. J., Annett, J. F., Souquet, J.-R. and Mackenzie, A. P. Spin-Orbit Coupling and ${\bm k}$-Dependent Zeeman Splitting in Strontium Ruthenate. {\em J. Phys.: Condens. Matter}, {\bf 2011}, {\em 23}, 094201. DOI: 10.1088/0953-8984/23/9/094201.
   \bibitem[Suderow(2009)]{Suderow2009}
   Suderow, H., Crespo, V. Guillamon, I., Vieira, S., Servant, F., Lejay, P., Brison, J. P. and Flouquet, J. A Nodeless Superconducting Gap in Sr$_2$RuO$_4$ from Tunneling Spectroscopy. {\em New J. Phys.} {\bf 2009}, {\em 11}, 093004. DOI: 10.1088/1367-2630/11/9/093004.
   \bibitem[Berry(1984)]{Berry1984}
   Berry, M. V. Quantal Phase Factors Accompanying Adiabatic Changes. {\em Proc. R. Soc. Lond. A} {\bf 1984}, {\em 392}, 45-57. DOI: 10.1098/rspa.1984.0023.
   \bibitem[Griffiths(2005)]{Griffiths2005}
   Griffiths, D. J. {\em Introduction to Quantum Mechanics}, Second Edition, Pearson, Upper Saddle River, NJ, 2005. ISBN: 978-1107179868.
 \bibitem[Hattori(2013)]{Hattori2013}
  Hattori, T., Karube, K., Ihara, Y., Ishida, K., Deguchi, K., Sato, N. K. and Yamamura, T. Spin Susceptibility in the Superconducting State of the Ferromagnetic Superconductor UCoGe. {\em Phys. Rev. B} {\bf 2013}, {\em 88}, 085127. DOI: 10.1103/PhysRevB.88.085127.
  \bibitem[Aoki(2012)]{Aoki2012}
  Aoki, D. and Flouquet J. Ferromagnetism and Superconductivity in Uranium Compounds. {\em J. Phys. Soc. Jpn.} {\bf 2012}, {\em 81}, 011003. DOI: 10.1143/JPSJ.81.011003.
  \bibitem[Gannon(2012)]{Gannon2012}
  Gannon, W. J., Halperin, W. P., Rastovski, C., Eskildsen, M. R., Dai, P. C. and Stunault, A. Magnetism in the Superconducting State of UPt$_3$ from Polarized Neutron Diffraction. {\em Phys. Rev. B} {\bf 2012}, {\em 86}, 104510.  DOI: 10.1103/PhysRevB.86.104510.
  \bibitem[Scharnberg(1985)]{Scharnberg1985}
  Scharnberg, K. and Klemm R. A. Upper Critical Field in $p$-Wave Superconductors with Broken Symmetry.  {\em Phys. Rev. Lett.} {\bf 1985}, {\em 54}, 2445-48.  DOI: 10.1103/PhysRevLett.54.2445.
 \bibitem[Hall(2016)]{Hall2016}
  Hall, B. E. and Klemm, R. A. Microscopic Model of the Knight Shift in Anisotropic and Correlated Metals. {\em J. Phys.: Condens. Matter} {\bf 2016}, {\em 28}, 03LT01 (14 pages). DOI: 10.1088/0953-8984/28/3/03LT01.
  \bibitem[Haug(2008)]{Haug2008}
  Haug, H. J. W. and Jauho, A.-P. {\em Quantum Kinetics in Transport and Optics of Semiconductors}; Springer, Berlin, Germany, 2008. ISBN 978-3-540-73561-8.
  \bibitem[Klemm(1980)]{Klemm1980}
   Klemm, R. A. and Clem, J. R. Lower Critical Field of an Anisotropic Type-II Superconductor.  {\em Phys. Rev. B} {\bf 1980}, {\em 21}, 1868-75. DOI: 10.1103/PhysRevB/21.1868.
  \bibitem[Klemm(1993)]{Klemm1993}
   Klemm, R. A. Lower Critical Field of a Superconductor with Uniaxial Anisotropy. {\em Phys. Rev. B} {\bf 1993}, {\em 47}, 14630. DOI: 10.1103/PhysRevB.47.14630
  \bibitem[Anderson(1961)]{Anderson1961}
  Anderson P. W. Localized Magnetic States in Metals. {\em Phys. Rev.} {\bf 1961}, {\em 124}, 41-53. DOI: 10.1103/PhysRev.124.41.
  \bibitem[Loerscher(2013)]{Loerscher2013}
   L{\"o}rscher, C., Zhang, J., Gu, Q. and Klemm, R. A. Anomalous Angular Dependence of the Upper Critical Inductiion of Orthorhombic Rerromagnetic Superconductors with Completely Broken $p$-Wave Symmetry. {\em Phys. Rev. B} {\bf 2013}, {\em 88},024504. DOI: 10.1103/PhysRevB.88.024504.
   \bibitem[Zhang(2014b)]{Zhang2014b}
   Zhang, J, L{\"o}rscher, C., Gu, Q. and Klemm, R. A. First-Order Chiral to Non-Chiral Transition in the Angular Dependence of the Upper Critical Induction of the Scharnberg-Klemm $p$-Wave Pair State. {\em J. Phys.: Condens. Matter} {\bf 2014}, {\em 26}, 252202. DOI: 10.1088-8984/26/25/252202.
   \bibitem[Rammer(1986)]{Rammer1986}
  Rammer, J. and Smith, H. Quantum Field-Theoretical Methods in Transport Theory of Metals. {\em Rev. Mod. Phys.} {\bf 1986}, {\em 58}, 323-359. DOI: 10.1103/RevModPhys.58.323
  \bibitem[Drozdov(2015)]{Drozdov2015}
  Drozdov, A. P., Eremets, M. I., Troyan, I. A., Ksenofontov, V. and Shylin, S. I.  Conventional Superconductivity at 203 K at High Pressures in the Sulfur Hydride System. {\em Nature}, {\bf 2015}, {\em 525}, 73-77. DOI: 10.1038/nature14964.
  \bibitem[Werthamer(1966)]{Werthamer1966}
  Werthamer, N. R., Helfand, E. and Hohenberg, P. C. Temperature and Purity Dependence of the Superconducting Critical Field $H_{c2}$. III. Spin and Spin-Orbit Effects. {\em Phys. Rev.} {\bf 1966}, {\em 147}, 295-302. DOI: 10.1103/PhysRev.147.295.
  \bibitem[Klemm(1975)]{Klemm1975}
  Klemm, R. A., Luther, A. and Beasley, M. R. Theory of the Upper Critical Field in Layered Superconductors.  {\em Phys. Rev. B} {\bf 1975}, {\em 12}, 877-91. DOI: 10.1103/PhysRevB.12.871.
  \bibitem[Abrikosov(1963)]{Abrikosov1963}
  Abrikosov, A. A., Gor'kov, L. P. and Dzyaloshinskii, I. E. {\em Methods of Quantum Field Theory in Statistical Physics} Prentice-Hall, Englewood Cliffs, NJ, 1963. ISBN: 978-0486632285.
  \bibitem[Gorkov(1959)]{Gorkov1959}
  Gor'kov, L. P. Microscopic Derivation of the Ginzburg-Landau Equations in the Theory of Superconductivity. {\em Sov. Phys. JETP}, {\bf 1959}, {\em 9}, 1364-67.
  \bibitem[Rickayzen(1969)]{Rickayzen1969}
  Rickayzen, G. The Theory of Bardeen, Cooper, and Schrieffer, in Parks, R. D. {\em Superconductivity}, Marcel Dekker, Inc., New York, 1969, pp. 51-115. ISBN: 978-0824715212.
  \bibitem[Klemm(2015)]{Klemm2015}
  Klemm R. A. Pristine and Intercalated Transition Metal Dichalcogenide Superconductors. {\em Physica C}, {\bf 2015}, {\em 514}, 86-94. DOI: 10.1016/j.physc.2015.02.023.
  \bibitem[Zhong(2016)]{Zhong2016}
  Zhong, Y., Wang, Y., Han, S., Lv, Y.-F., Wang, W.-L., Zhang, D., Ding, H., Zhang, Y.-M., Wang, L., He, K., Zhong, R., Schneeloch, A. A., Gu, G.-D., Song, C.-L., Ma, X.-C. and Xue, Q.-K. Nodeless Pairing in Superconducting Copper-Oxide Monolayer Films on Bi$_2$Sr$_2$CaCu$_2$O$_{8+\delta}$. {\em Sci. Bull.} {\bf 2016}, {\em 61}, 1239-1247. DOI: 10.1007/s11434-016-1145-4.
  \bibitem[Li(1999)]{Li1999}
  Li, Q., Tsay, Y. N., Suenaga, M., Klemm, R. A., Gu, G. D. and Koshizuka, N. Bi$_2$Sr$_2$CaCu$_2$O$_{8+\delta}$ Bicrystal $c$-Axis Twist Josephson Junctions:  A New Phase-Sensitive Test of Order Parameter Symmetry. {\em Phys. Rev. Lett.}, {\bf 1999}, {\em 83}, 4160-4163.  DOI: 10.1103/PhysRevLett/83.4160.

\end{thebibliography}
\end{document}